%% file: main.tex
\documentclass[conference,compsoc]{IEEEtran}

\newif\ifpublicversion
\ifdefined\submissionversion
  \publicversionfalse
\else
  \publicversiontrue
\fi

\ifCLASSOPTIONcompsoc
  \usepackage[nocompress]{cite}
\else
  \usepackage{cite}
\fi
\usepackage{tikz}
\usetikzlibrary{arrows.meta,positioning,calc,shapes.geometric,shadows.blur,backgrounds}
\usepackage{pgfplots}
\pgfplotsset{compat=1.18}
\usepgfplotslibrary{groupplots}
\usepackage{amsmath}
\usepackage{amssymb}
\usepackage{algorithm}
\usepackage{algpseudocode}
\algrenewcommand{\algorithmiccomment}[1]{\hfill{\footnotesize$\triangleright$~#1}}
\usepackage{booktabs}
\usepackage{tabularx}
\usepackage{array}
\usepackage{ragged2e}
\usepackage{multirow}
\usepackage{graphicx}
\usepackage{afterpage}
\usepackage{float}
\usepackage{cuted}
\usepackage{xcolor}
\usepackage{url}
\usepackage[hidelinks]{hyperref}
\usepackage{listings}
\usepackage{fontawesome5}

\definecolor{insightblue}{HTML}{1F4D63}
\definecolor{insightfill}{HTML}{FFFFFF}
\definecolor{insightSynthTint}{HTML}{D6E4F0}
\definecolor{insightSynthEdge}{HTML}{2F5E8A}
\definecolor{insightDetectTint}{HTML}{EFE3C4}
\definecolor{insightDetectEdge}{HTML}{8A6A2F}
\definecolor{insightValidTint}{HTML}{C9DFD0}
\definecolor{insightValidEdge}{HTML}{2D6A4F}
\definecolor{insightaccent}{HTML}{B85C2A}
\definecolor{algframe}{HTML}{2F5E8A}
\definecolor{alghead}{HTML}{EAF2F8}
\definecolor{algline}{HTML}{D8E2EA}

\lstdefinestyle{cmotiv}{%
  language=C,
  basicstyle=\ttfamily\footnotesize,
  keywordstyle=\color{blue!75!black}\bfseries,
  commentstyle=\color{green!45!black}\itshape,
  stringstyle=\color{red!70!black},
  numbers=left,
  numberstyle=\footnotesize\color{black!55},
  numbersep=6pt,
  frame=single,
  framesep=4pt,
  framerule=0.4pt,
  xleftmargin=14pt,
  xrightmargin=2pt,
  showstringspaces=false,
  breaklines=true,
  columns=fullflexible,
  keepspaces=true,
  morekeywords={size_t,ssize_t,argc,argv,getenv,setenv,xstrdup,obstack_grow,obstack_finish,expand_line,argcv_get,execv,strlen,strcspn},
  literate={->}{{$\rightarrow$}}1
}

\lstdefinestyle{candidate}{%
  basicstyle=\ttfamily\footnotesize,
  frame=single,
  framesep=4pt,
  framerule=0.4pt,
  xleftmargin=4pt,
  xrightmargin=2pt,
  showstringspaces=false,
  breaklines=true,
  columns=fullflexible,
  keepspaces=true,
  escapeinside={(*@}{@*)},
  literate={->}{{$\rightarrow$}}1
}

\lstdefinestyle{prompt}{%
  style=candidate,
  basicstyle=\ttfamily\scriptsize,
  framesep=2pt,
  xleftmargin=1pt,
  xrightmargin=1pt,
  breakautoindent=false,
  breakindent=0pt,
  aboveskip=3pt,
  belowskip=0pt
}

\ifpublicversion
\newcommand{\submittedtext}{This work has been submitted to the IEEE for possible publication. Copyright may be transferred without notice, after which this version may no longer be accessible.}
\newcommand{\submittednotice}{%
  \begin{tikzpicture}[remember picture,overlay]
    \node[
      anchor=south,
      align=center,
      font=\sffamily\scriptsize,
      text=black!70,
      text width=0.94\paperwidth
    ] at ([yshift=0.34in]current page.south) {\submittedtext};
  \end{tikzpicture}%
}
\AddToHook{shipout/foreground}{%
  \ifnum\value{page}=1\relax
    \submittednotice
  \fi
}
\fi

\newcolumntype{Y}{>{\RaggedRight\arraybackslash}X}
\newcolumntype{Z}{>{\Centering\arraybackslash}X}
\newcolumntype{L}[1]{>{\RaggedRight\arraybackslash}p{#1}}
\newcolumntype{C}[1]{>{\Centering\arraybackslash}p{#1}}

\newcommand{\fullsupport}{\tikz[baseline=-0.55ex]\fill[black!85] (0,0) circle (0.55ex);}
\newcommand{\partialsupport}{\tikz[baseline=-0.55ex]{\draw[black!70, line width=0.25pt] (0,0) circle (0.55ex);\fill[black!85] (90:0.55ex) arc (90:270:0.55ex) -- cycle;}}
\newcommand{\nosupport}{\tikz[baseline=-0.55ex]\draw[black!70, line width=0.35pt] (0,0) circle (0.55ex);}

\newcommand{\code}[1]{\texttt{#1}}

\newif\ifblind
\ifpublicversion
  \blindfalse
\else
  \blindtrue
\fi

\newcommand{\sysname}{Antiproof}
\newcommand{\enginename}{Antigraph}
\newcommand{\ecpg}{eCPG}

\newcommand{\cve}[2]{%
  \href{https://cve.mitre.org/cgi-bin/cvename.cgi?name=CVE-#1-#2}{\mbox{\texttt{CVE-#1-#2}}}%
}

\newcommand{\numdetectors}{32}
\newcommand{\numcoveredclasses}{45}

\newcommand{\numlanguages}{8}
\newcommand{\numprojects}{50}
\newcommand{\numcves}{12}
\newcommand{\numduplicates}{three}
\newcommand{\numcandidates}{17{,}574}
\newcommand{\numfindings}{1{,}212}
\newcommand{\numpoes}{510}
\newcommand{\nummanualverified}{100}
\newcommand{\detspeedup}{25}
\newcommand{\detbatchmin}{2.3}
\newcommand{\joernbatchmin}{59}
\newcommand{\numscancost}{2{,}828}
\newcommand{\bballtotal}{46}
\newcommand{\bbsupported}{46}
\newcommand{\bbdetected}{44}
\newcommand{\bbdetectrate}{96}
\newcommand{\bbsota}{20}
\newcommand{\kevtotal}{20}
\newcommand{\kevdetected}{20}
\newcommand{\numtargets}{66}
\newcommand{\numdetected}{64}

\newif\ifcomments
\commentstrue
\ifcomments
    \providecommand{\ion}[1]{{\color{orange}{/* ion: #1 */}}}
    \providecommand{\agent}[1]{{\color{purple}{/* agent: #1 */}}}
\else
    \providecommand{\ion}[1]{}
    \providecommand{\agent}[1]{}
\fi

\begin{document}

\date{}

\title{\sysname{}: Synthesizing Vulnerability Detectors and Proofs of Exploitability}
\ifpublicversion
  \author{%
    \IEEEauthorblockN{Alon Shakevsky}
    \IEEEauthorblockA{UC Berkeley\\shakevsky@berkeley.edu}
    \and
    \IEEEauthorblockN{Corban Villa}
    \IEEEauthorblockA{UC Berkeley}
    \and
    \IEEEauthorblockN{Ion Stoica}
    \IEEEauthorblockA{UC Berkeley}
    \and
    \IEEEauthorblockN{Raluca Ada Popa}
    \IEEEauthorblockA{UC Berkeley}
  }%
\else
  \author{\mbox{}}%
\fi

\maketitle

\begin{abstract}
Discovering vulnerabilities before attackers exploit them requires high recall and reliable automatic validation, but existing approaches struggle to achieve both without prohibitive cost. We present \sysname{}, an end-to-end vulnerability discovery system that combines \emph{neuro-symbolic detector synthesis} for high-recall discovery with \emph{proof-of-exploitability oracles} for automatic validation. \sysname{} learns and iteratively refines static detectors from vulnerability datasets, then validates candidates by verifying whether executable proofs demonstrate concrete attacker capabilities. Evaluated on BountyBench and our curated KEVBench dataset, \sysname{} detects 64 of 66 vulnerabilities, improving recall by more than 60 percentage points over static-analysis and neuro-symbolic baselines. In a scan of \numprojects{} widely deployed systems, \sysname{} uncovered several hundred previously unknown vulnerabilities. We are responsibly disclosing all confirmed zero-days and have received \numcves{} CVE assignments to date, including remote code execution vulnerabilities in Ray, SGLang, vLLM, and LiteLLM that could allow attackers to take over LLM training and inference systems.
\end{abstract}

\IEEEpeerreviewmaketitle

\section{Introduction}
\label{sec:intro}

Exploitable vulnerabilities are the most common initial-access vector in organizational data breaches~\cite{mandiant2025mtrends,verizon2026dbir} and the leading root cause of ransomware incidents~\cite{sophos2025stateofransomware}. Because the general problem of discovering such flaws is undecidable~\cite{rice1953classes}, practical systems must approximate and trade off \textit{recall}, \textit{precision}, and \textit{cost}. Addressing this tradeoff has become urgent as exploitation timelines have shrunk from years to hours~\cite{zerodayclock2026,sonicwall2025cyberthreatreport}, leaving defenders less time to discover and remediate vulnerabilities. An ideal solution for discovering exploitable flaws before attackers exploit them should provide high recall, reliable automatic validation (high precision), and low cost.

Automated vulnerability discovery has been extensively studied, but no existing method achieves these properties simultaneously across a broad range of vulnerability classes~\cite{zhang2025llms,rohlf2025lifecycle}. Static analysis tools can scan large systems~\cite{ayewah2008static,bessey2010few}, but require experts to specify vulnerability patterns and manually review findings. These tools often prioritize precision over recall to reduce false positives and review overhead~\cite{bessey2010few}. Dynamic approaches, such as fuzzing and symbolic execution, are effective in finding memory-safety vulnerabilities, but do not easily generalize to vulnerability classes that lack reliable bug oracles. These approaches often become prohibitively expensive for scanning large systems due to path explosion and deep program states~\cite{cha2012mayhem,baldoni2018survey,li2018fuzzing,manes2019art}.

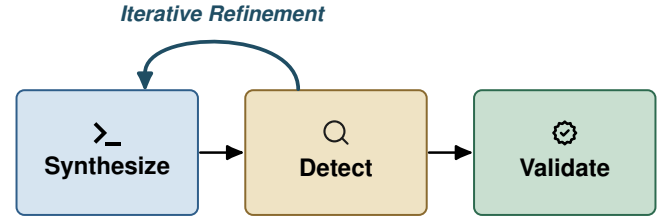
\begin{figure}[!t]
\centering
\begin{tikzpicture}[
    >=Latex,
    every node/.style={font=\sffamily},
    stage/.style={
        rounded corners=3pt,
        draw=black,
        line width=0.75pt,
        fill=insightfill,
        inner sep=1pt,
        align=center,
        minimum width=24mm,
        minimum height=16.5mm
    },
    stageSynth/.style={
        stage, draw=insightSynthEdge, fill=insightSynthTint
    },
    stageDetect/.style={
        stage, draw=insightDetectEdge, fill=insightDetectTint
    },
    stageValid/.style={
        stage, draw=insightValidEdge, fill=insightValidTint
    },
    stageicon/.style={
        inner sep=0pt
    },
    stagelabel/.style={
        font=\sffamily\bfseries\small,
        text=black
    },
    flow/.style={
        -{Latex[length=2.7mm,width=2.3mm,round]},
        line width=0.95pt,
        draw=black,
        line cap=round,
        shorten <=0.5pt, shorten >=0.5pt
    },
    refinearrow/.style={
        -{Latex[length=3mm,width=2.6mm,round]},
        line width=1.4pt,
        draw=insightblue,
        line cap=round
    }
]

\node[stageSynth] (s1) at (0,0) {};
\node[stageicon] at ([yshift=2.6mm]s1.center)
    {\includegraphics[height=4mm]{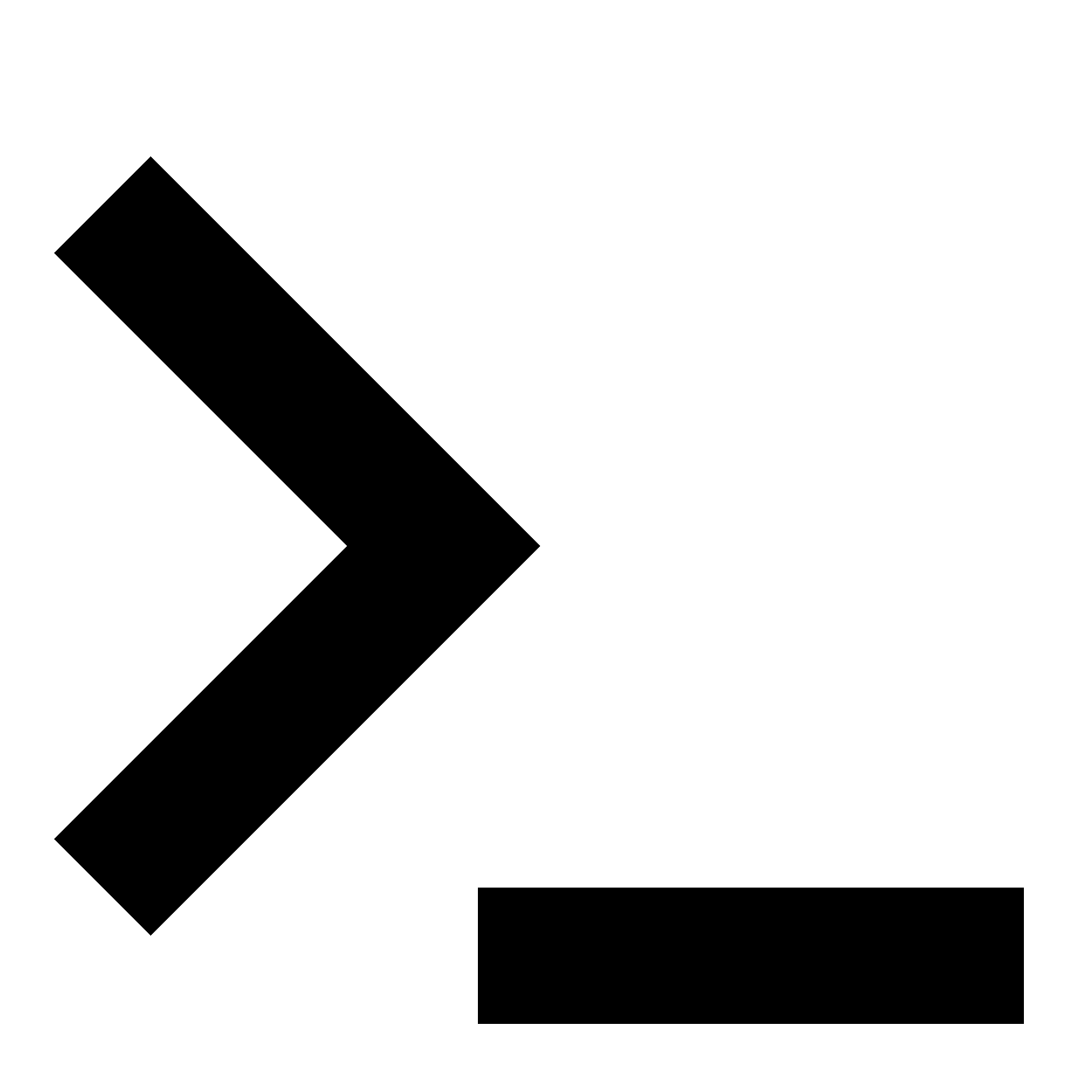}};
\node[stagelabel] at ([yshift=-2.0mm]s1.center) {Synthesize};

\node[stageDetect, right=6mm of s1] (s2) {};
\node[stageicon] at ([yshift=2.6mm]s2.center)
    {\includegraphics[height=4mm]{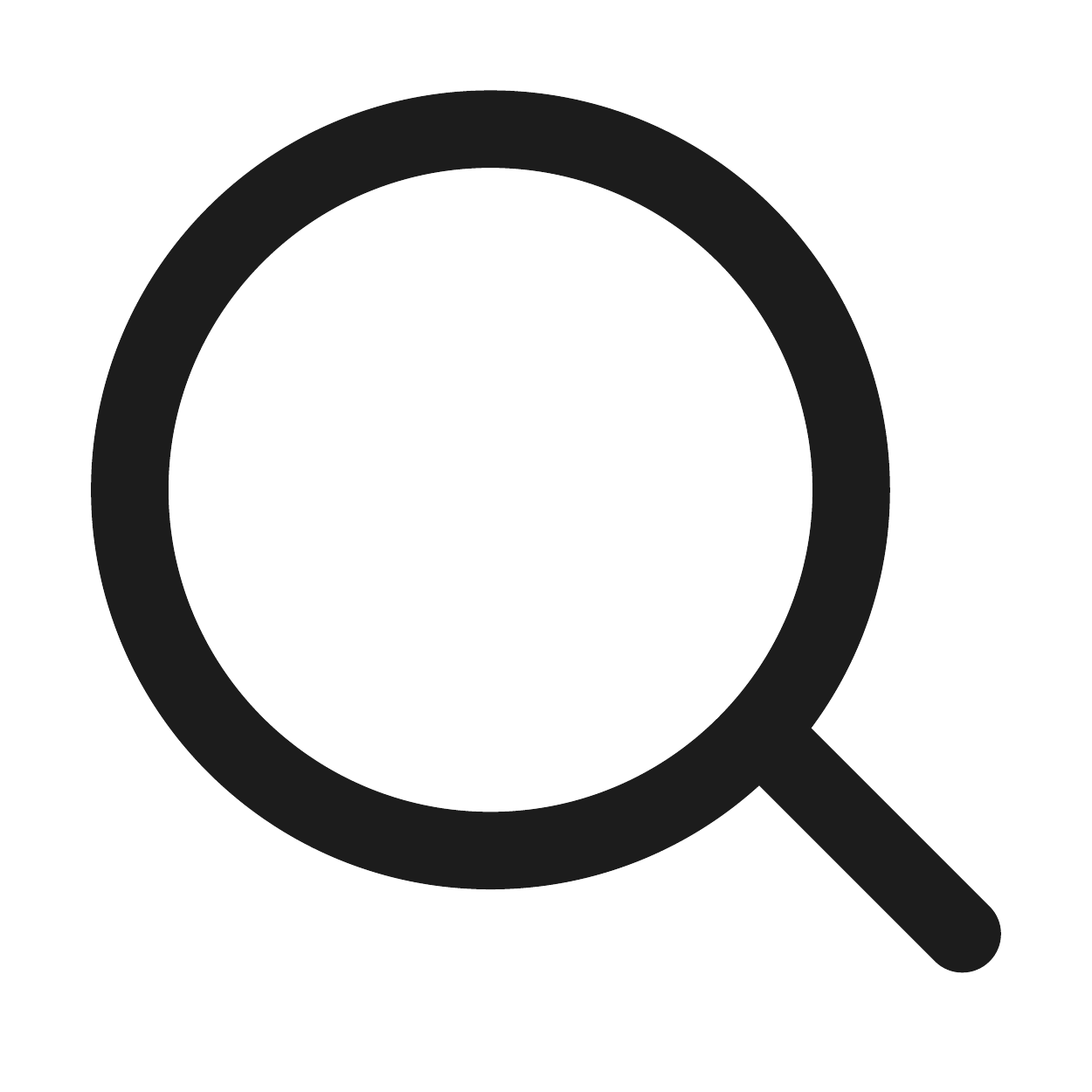}};
\node[stagelabel] at ([yshift=-2.0mm]s2.center) {Detect};

\node[stageValid, right=6mm of s2] (s3) {};
\node[stageicon] at ([yshift=2.6mm]s3.center)
    {\includegraphics[height=4mm]{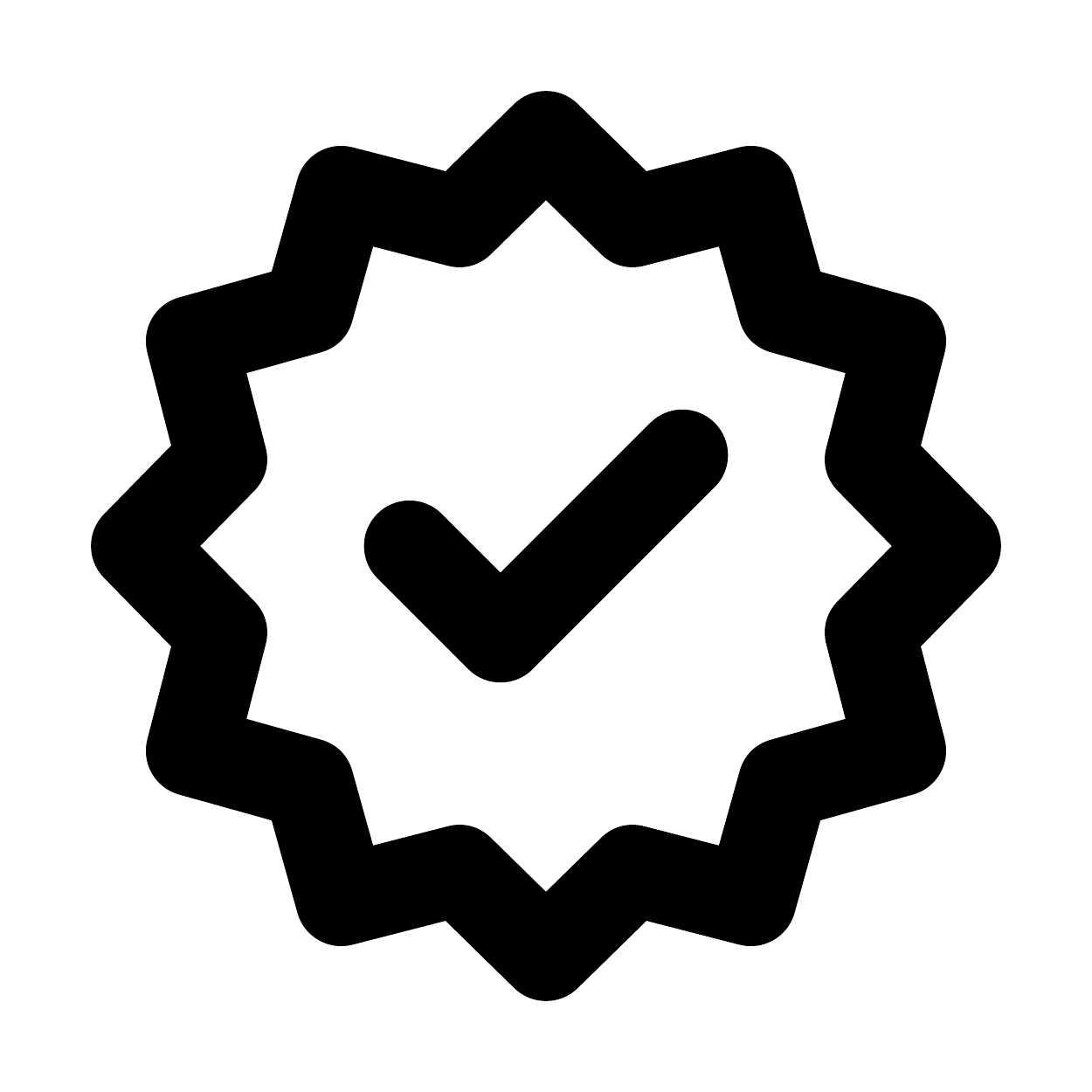}};
\node[stagelabel] at ([yshift=-2.0mm]s3.center) {Validate};

\draw[flow] (s1.east) -- (s2.west);
\draw[flow] (s2.east) -- (s3.west);

\coordinate (mid)    at ($(s1.north)!0.5!(s2.north)$);
\coordinate (rstart) at ($(s2.north) + (-5mm, 0)$);
\coordinate (rend)   at ($(s1.north) + (5mm, 0)$);
\draw[refinearrow]
    (rstart)
    .. controls ($(rstart) + (0, 8mm)$)
            and ($(rend)   + (0, 8mm)$)
    .. (rend);
\node[font=\sffamily\bfseries\footnotesize\itshape, text=insightblue, anchor=south]
    at ($(mid) + (0, 7.8mm)$)
    {Iterative Refinement};
\end{tikzpicture}%
\caption{\sysname{} synthesizes and iteratively refines high-recall static detectors from vulnerability datasets, scans target repositories for vulnerability candidates, and validates them with executable proofs of exploitability.}
\label{fig:insight}
\end{figure}

\begin{table*}[t]
\caption{Qualitative comparison of vulnerability discovery and validation systems.}
\label{tab:approaches}
\centering
\scriptsize
\renewcommand{\arraystretch}{1.12}
\setlength{\tabcolsep}{1.1pt}
\begin{tabular}{@{}L{0.16\textwidth}L{0.341\textwidth}C{0.058\textwidth}C{0.075\textwidth}C{0.052\textwidth}C{0.072\textwidth}C{0.083\textwidth}C{0.052\textwidth}@{}}
\toprule
\textbf{Approach} & \textbf{Representative systems} & \multicolumn{3}{c}{\textbf{Discovery}} & \multicolumn{3}{c}{\textbf{Validation}} \\
\cmidrule(lr){3-5}\cmidrule(lr){6-8}
& & \shortstack{\textbf{Learned}\\\textbf{patterns}} & \shortstack{\textbf{Class}\\\textbf{coverage}} & \shortstack{\textbf{Low}\\\textbf{cost}} & \shortstack{\textbf{Executable}\\\textbf{evidence}} & \shortstack{\textbf{Exploitability}\\\textbf{oracle}} & \shortstack{\textbf{Low}\\\textbf{cost}} \\
\midrule
Static analysis & CodeQL~\cite{avgustinov2016ql}, Semgrep~\cite{semgrep_docs} & \nosupport & \fullsupport & \fullsupport & \nosupport & \nosupport & \nosupport \\
Neuro-symbolic detectors & IRIS~\cite{li2025iris}, QLCoder~\cite{wang2025qlcoder}, KNighter~\cite{yang2025knighter}, MoCQ~\cite{li2025automated} & \fullsupport & \partialsupport & \fullsupport & \nosupport & \nosupport & \nosupport \\
LLM-guided auditing & RepoAudit~\cite{guo2025repoaudit}, Co-RedTeam~\cite{he2026co} & \nosupport & \partialsupport & \nosupport & \partialsupport & \nosupport & \nosupport \\
LLM security agents & Claude Mythos\textsuperscript{*}~\cite{anthropic_mythos_preview_2026}, Codex Security\textsuperscript{*}~\cite{openai_codex_security_2026} & \nosupport & \fullsupport & \nosupport & \fullsupport & \nosupport & \nosupport \\
Fuzzing\textsuperscript{**} & AFL++~\cite{fioraldi2020afl++}, OSS-Fuzz~\cite{google_oss-fuzz_2025}, Syzkaller~\cite{google_syzkaller_2025} & \nosupport & \partialsupport & \nosupport & \fullsupport & \partialsupport & \partialsupport \\
Symbolic execution\textsuperscript{**} & Mayhem~\cite{cha2012mayhem}, Arbiter~\cite{vadayath2022arbiter}, SAILOR~\cite{shafiuzzaman2026sailor} & \nosupport & \partialsupport & \nosupport & \fullsupport & \partialsupport & \nosupport \\
PoC generation & AnyPoC~\cite{zhao2026anypoc}, FaultLine~\cite{nitin2025faultline}, \mbox{CVE-GENIE}~\cite{ullah2025cve} & \nosupport & \nosupport & \nosupport & \fullsupport & \partialsupport & \partialsupport \\
\textbf{\sysname{}} & \textbf{This work} & \fullsupport & \fullsupport & \fullsupport & \fullsupport & \fullsupport & \fullsupport \\
\bottomrule
\end{tabular}
\vspace{0.15em}
\begin{flushleft}
\fullsupport{}, \partialsupport{}, and \nosupport{} denote full, partial, and no support, respectively. \textsuperscript{*}We do not evaluate proprietary systems because we do not have access to their source code, and we use public reports for comparison. \textsuperscript{**}Fuzzing and symbolic execution are limited to specific vulnerability classes, primarily memory safety.
\end{flushleft}
\end{table*}

Large language models (LLMs) have shown strong code-reasoning capabilities that can help address this challenge by automating parts of vulnerability discovery, exploitation, and patching~\cite{zhang2025llms,sheng2025llms,zhou2025large,heelan2026industrialisation}. Proprietary LLM agents, such as Claude Mythos~\cite{anthropic_mythos_preview_2026,anthropic_glasswing_2026} and Codex Security~\cite{openai_codex_security_2026,openai2025aardvark}, demonstrate that LLM-based systems can uncover thousands of vulnerabilities in production software. Despite this promise, LLM-based vulnerability discovery remains difficult to scale because LLM performance relies on careful context management~\cite{mei2025survey}, and a single project scan can require hundreds of millions of tokens and multiple days~\cite{li2026llm}. Continuous scanning is therefore impractical on large systems such as the Linux kernel. Moreover, high false-positive rates from LLM hallucinations and reward hacking create a validation bottleneck that requires expensive expert review to determine whether a finding is an exploitable vulnerability or a false positive~\cite{li2026llm,kalai2025language,du2026reducing}. For example, Anthropic reports running Claude Mythos roughly a thousand times on OpenBSD at a total cost of USD~20{,}000 and hiring security experts to manually validate its reports before disclosure~\cite{anthropic_mythos_preview_2026}. Recent work combines LLMs with static analysis~\cite{li2025automated,yang2025knighter}, symbolic execution~\cite{shafiuzzaman2026sailor}, and proof-of-concept generation~\cite{zhao2026anypoc,nitin2025faultline,ullah2025cve}, but remains limited to specific vulnerability classes or still
requires manual expert review to validate exploitability. As shown in Table~\ref{tab:approaches}, these limitations motivate an end-to-end system that achieves high recall, automatic executable validation, and low cost.

In this work, we propose \sysname{}, an end-to-end system that provides all three properties for a wide range of vulnerabilities. Our key insight, as shown in Figure~\ref{fig:insight}, is to combine high-recall discovery with proofs of exploitability. To achieve this, we contribute two techniques: (1) \sysname{} uses LLMs to synthesize reusable detectors and iteratively refine them for high-recall detection on a dataset of known vulnerabilities. To find vulnerabilities that data-flow analysis tools miss, including semantic vulnerabilities, we introduce \ecpg{} traversals. An \ecpg{} is a graph representation of code that extends the code property graph~\cite{yamaguchi2014modeling} with detector-specific virtual nodes and edges to support modeling new vulnerability classes. (2) \sysname{} uses LLMs to create validation environments to validate the detected vulnerability candidates and generate proofs that demonstrate a concrete attacker capability, such as code execution, file access, SSRF, or authentication bypass. \sysname{} implements proof-of-exploitability oracles that verify the proofs in a challenge-response protocol that rejects LLM hallucinations. This reduces vulnerability validation to proof search under a deterministic oracle.

Building a practical system from this insight raises three challenges:

\begin{enumerate}
    \item \textbf{Detectors that generalize.} Synthesized detectors must capture the underlying vulnerability pattern rather than overfit to a specific vulnerability or project. Overly specific detectors may miss variants of the vulnerability class (low recall), while overly broad detectors produce too many candidates for validation (high validation cost). The system must generalize across project sizes, programming languages, frameworks, and vulnerability classes, including logic and semantic flaws that conventional static analysis struggles to express.
    \item \textbf{Project-scale discovery.} Large systems with millions of lines of code, such as the Linux kernel, do not fit in a single LLM context window. The system must have high code coverage and construct only the minimal context that an LLM agent requires to validate each vulnerability candidate while avoiding prohibitive LLM inference cost.
    \item \textbf{Exploitability validation.} LLMs can generate plausible but incorrect vulnerability reports and create a validation bottleneck that requires experts to manually review the reports. Automatic validation requires oracles that execute against the target system and check whether the vulnerability candidate enables an attacker-controlled security violation. These oracles must reject proofs that do not demonstrate practical risk to scale vulnerability validation.
\end{enumerate}

\begin{figure*}[!t]
\centering
\definecolor{legReach}{HTML}{1D4ED8}
\definecolor{legRed}{HTML}{B91C1C}
\definecolor{legPink}{HTML}{FEE2E2}
\definecolor{legInk}{HTML}{0F172A}
\begin{tikzpicture}[
    x=1cm, y=1cm,
    font=\footnotesize\sffamily,
    every node/.style={font=\footnotesize\sffamily, text=legInk},
    reach/.style={-{Latex[length=2.5mm, width=2mm]}, line width=1.4pt, draw=legReach},
    bridge/.style={-{Latex[length=2.5mm, width=2mm]}, line width=1.4pt, draw=legReach,
        dash pattern=on 1pt off 2pt},
    cand/.style={rectangle, rounded corners=2pt, draw=legRed, fill=white,
        line width=1.5pt, minimum width=0.85cm, minimum height=0.32cm, inner sep=0pt},
    sinkbox/.style={rectangle, rounded corners=2pt, draw=legRed, fill=legPink,
        line width=1.2pt, minimum width=0.85cm, minimum height=0.32cm, inner sep=0pt}
]
\draw[reach] (0,0) -- (0.7,0);
\node[anchor=west] at (0.85,0) {\ecpg{} path};
\draw[bridge] (4.8,0) -- (5.5,0);
\node[anchor=west] at (5.65,0) {\ecpg{} virtual edge};
\node[cand] at (10.9,0) {};
\node[anchor=west] at (11.45,0) {candidate};
\node[sinkbox] at (13.6,0) {};
\node[anchor=west] at (14.15,0) {sink};
\end{tikzpicture}

\vspace{0.7ex}

\def\paneltitle#1{%
  {\centering\sffamily\bfseries\footnotesize\textcolor{legInk}{#1}\par}%
  \vspace{2.5pt}%
  {\color{legInk}\hrule height 0.9pt}%
  \vspace{4pt}}

\begin{minipage}[t]{0.32\linewidth}
\paneltitle{Vulnerable code}
\begin{lstlisting}[style=cmotiv, basicstyle=\ttfamily\scriptsize]
// attacker-controlled (state.c):
setenv("USER", value, 1);

// expand_line (utility.c):
p = getenv("USER");
obstack_grow(&obs, p, strlen(p));
...
cmd = obstack_finish(&obs);

// start_login (pty.c):
argcv_get(cmd, "", &argc, &argv);
execv(argv[0], argv);
\end{lstlisting}
\end{minipage}\hfill
\begin{minipage}[t]{0.32\linewidth}
\paneltitle{\ecpg{} traversal}
\centering
\resizebox{\linewidth}{!}{\input{figures/inetutils_traversal}}
\end{minipage}\hfill
\begin{minipage}[t]{0.32\linewidth}
\paneltitle{Candidate context}
\begin{lstlisting}[style=candidate, basicstyle=\ttfamily\scriptsize]
(*@\textcolor{insightblue}{\bfseries\#\#\# Enclosing function}@*)
// telnetd/pty.c:112-138
void start_login(...) {
  ...
  argcv_get(cmd, "", &argc, &argv);
  execv(argv[0], argv);
}

(*@\textcolor{insightblue}{\bfseries\#\#\# Static analysis observations}@*)
- sink: command_injection
  at pty.c:135
- env writer: setenv
  at state.c:1495, 1514
- env reader: getenv("USER")
  at utility.c:1740
- attack surface: main reaches
  start_login via telnetd_setup
\end{lstlisting}
\end{minipage}
\caption{Motivating example (simplified): \cve{2026}{24061} in GNU InetUtils \code{telnetd} allowed unauthenticated attackers to gain root access to over 212{,}000 internet-exposed devices. The root cause is that a remote attacker can set the \code{USER} environment variable, which \code{telnetd} injects unsanitized as an argument to the \code{login} program. \textbf{Left:} the path from the attacker-controlled \code{setenv} to the \code{execv} sink, which traditional data-flow analysis (CodeQL, Joern) misses without a propagation model for the custom allocator (\code{obstack}). \textbf{Middle:} \ecpg{} detects the path from attacker-controlled input (\code{net\_read}) to a dangerous sink (\code{execv}) and adds an observation about the \code{setenv} write and \code{getenv} read via a virtual edge. \textbf{Right:} the context \sysname{} prepares for the vulnerability candidate.}
\label{fig:motivation}
\end{figure*}
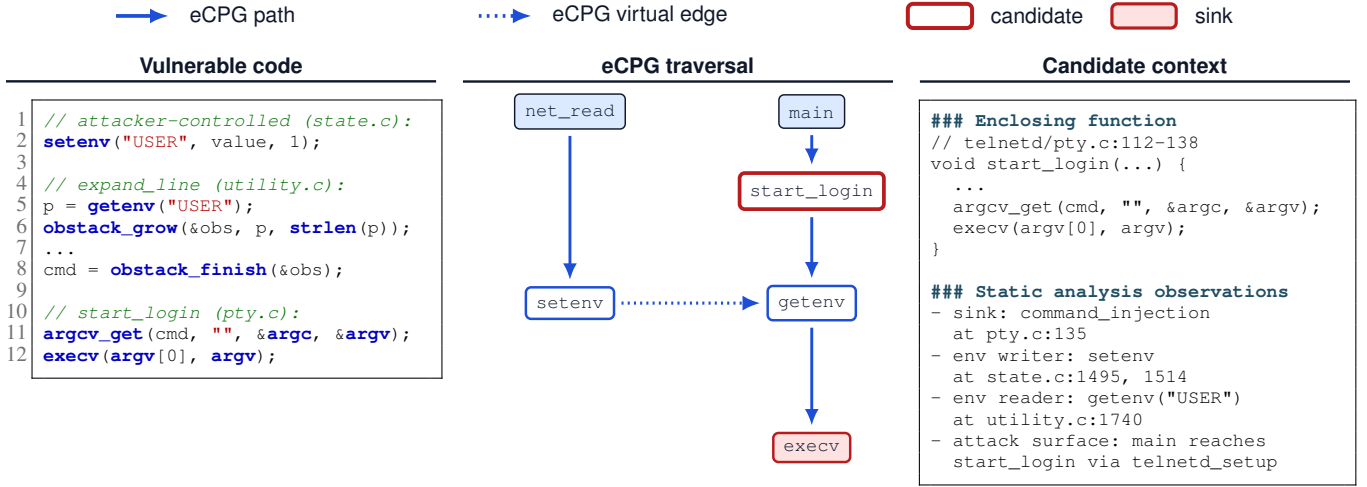

\sysname{} addresses these challenges and manages the inherent tradeoff between recall, precision, and cost. The system refines vulnerability detectors for high recall rather than precision against a vulnerability dataset. \sysname{} uses proofs of exploitability to shift the manual review bottleneck from every vulnerability candidate to a small set of validation environments and oracles. To run the synthesized detectors at scale, we implement \enginename{}, a native engine that constructs the \ecpg{} and evaluates detector traversals an order of magnitude faster than existing CPG tools (see Section~\ref{sec:efficiency}), which allows \sysname{} to continuously scan large systems without expensive LLM inference. This design aims to improve recall as the vulnerability dataset covers more patterns and as LLMs become more capable.

We evaluate \sysname{} on known-vulnerability benchmarks and real-world zero-day discovery. We curate KEVBench, a dataset of CISA Known Exploited Vulnerabilities~\cite{cisa_kev} in 2025 and 2026. \sysname{} detects \bbdetected{} of \bballtotal{} BountyBench~\cite{zhang2026bountybench} vulnerabilities and \kevdetected{} of \kevtotal{} KEVBench vulnerabilities, outperforming evaluated baselines by over 60 percentage points in recall. Across \numprojects{} widely deployed systems, \sysname{} generated \numpoes{} proofs of exploitability. Due to time constraints, we selected \nummanualverified{} findings for manual review and coordinated disclosure. We have received \numcves{} CVE assignments to date and discovered \numduplicates{} vulnerabilities before their CVEs were publicly disclosed. \sysname{} uncovered critical and high-severity flaws, including remote code execution vulnerabilities in Ray, SGLang, vLLM, and LiteLLM that allow attackers to take over LLM training and inference systems, steal GPU resources~\cite{oligo2024shadowray}, and poison production LLM models.

\section{Background}
\label{sec:background}

\subsection{Motivating Example}
\label{sec:bg-motiv}

Consider \cve{2026}{24061}, an authentication bypass vulnerability in GNU InetUtils \code{telnetd}~\cite{josefsson2026telnetd} from CISA's Known Exploited Vulnerabilities catalog~\cite{cisa_kev} that allowed attackers to gain unauthenticated root access to over 212{,}000 internet-exposed devices. The root cause is that \code{telnetd} passed unsanitized user-controlled input to the \code{login} program and an attacker could inject an argument that skips password verification, as shown in Figure~\ref{fig:motivation}.

Leading static analysis tools such as CodeQL~\cite{avgustinov2016ql} and Joern~\cite{yamaguchi2014modeling} miss this vulnerability without a propagation model for the custom allocator (\code{obstack\_grow}) that GNU InetUtils uses, which hides the flow between the source (\code{getenv}) and the sink (\code{execv}). Even with a complete model specified by an engineer or an LLM, taint analysis cannot verify whether the input is sufficiently sanitized. For example, \cve{2026}{28372}, disclosed four weeks later, exploits the insufficient sanitization in the patch to \cve{2026}{24061}. Since this vulnerability is not a memory-safety bug, approaches that rely on memory sanitizers~\cite{stepanov2015memorysanitizer,serebryany2012addresssanitizer,song2019sok}, including fuzzing, are not effective in detecting it. This gap motivates our approach, which combines high-recall static detection with executable validation to discover and confirm such critical vulnerabilities.

\subsection{Threat Model}
\label{sec:threat}

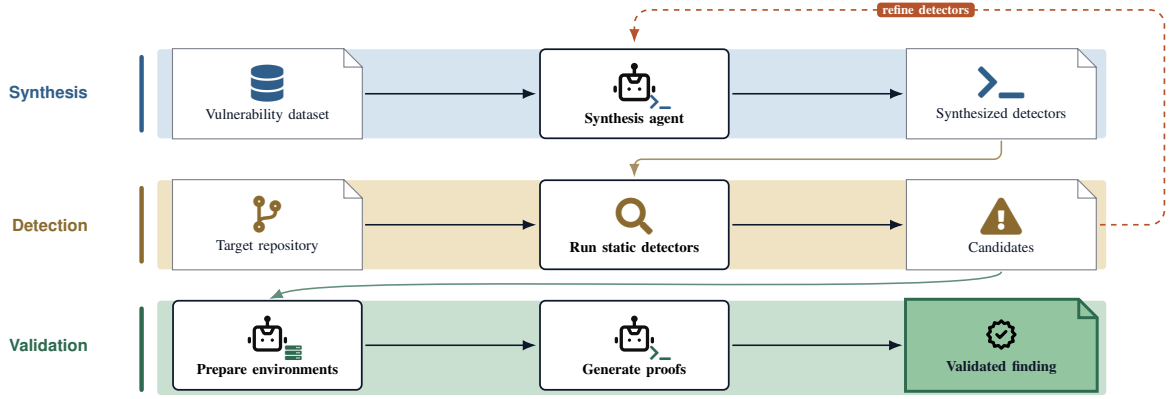
\begin{figure*}[!t]
\centering
\resizebox{\textwidth}{!}{\input{figures/architecture}}
\caption{\sysname{} system architecture. \textit{Synthesis} learns high-recall heuristics from a vulnerability dataset and refines them from observed false positives and false negatives. \textit{Detection} runs the detectors on a target repository to find vulnerability candidates. \textit{Validation} prepares a reusable PoE environment and a candidate-specific PoV environment, generates proofs, and checks each proof with its environment's oracle.}
\label{fig:architecture}
\end{figure*}

We define the \textit{attacker} as an untrusted actor interacting with the target system across a deployment trust boundary. As the systems under evaluation do not have precise definitions of their threat model, we use LLM agents to infer deployment configurations and trust boundaries from official documentation. We assume that the attacker controls inputs that a realistic deployment receives across a trust boundary, such as network requests, RPC messages, and serialized objects. We focus on systems where a validation environment can demonstrate exploitability using Docker containers, QEMU~\cite{bellard2005qemu}, or virtual machines.

We define an \textit{exploitable vulnerability} as a software flaw that enables an attacker to gain a new capability on a deployed system or on a user's system, such as code execution, file read, file write, authentication bypass, or denial of service. We do not consider flaws that do not enable new capabilities or are not reachable by the attacker as exploitable.

We define a vulnerability as \emph{known}, or \emph{one-day}, if it is publicly tracked with a CVE or explicitly acknowledged in an official advisory or project documentation. We define a vulnerability as \emph{previously unknown}, or \emph{zero-day}, if it affects the latest release at the time of discovery and was not publicly documented when \sysname{} found it.

\subsection{Code Property Graphs}
\label{sec:bg-cpg}

A \textit{code property graph} (CPG) is a graph representation of code that combines the abstract syntax tree, control-flow graph, and data-dependence graph into a single graph. CPGs have shown promise for modeling security vulnerabilities~\cite{yamaguchi2014modeling} with graph traversals~\cite{rodriguez2010graph}. Joern is an open-source tool that analyzes code through CPG queries and supports languages such as C/C++, Java, Python, JavaScript, PHP, and Go. In this work, we introduce extended CPG (\ecpg{}) traversals to model vulnerabilities that are missed by data-flow analysis tools, such as CodeQL and Joern. \sysname{} uses LLMs to generate and refine \ecpg{} queries that detect a wide range of vulnerability classes.

\subsection{Proofs of Vulnerability and Exploitability}
\label{sec:bg-pov-poe}

We define a \textit{proof of vulnerability} (PoV) as an executable test that uses a bug oracle, such as a test harness with assertions, to demonstrate a violation of a security property in isolation. The most widely deployed bug oracles are memory sanitizers~\cite{serebryany2012addresssanitizer,stepanov2015memorysanitizer,song2019sok}, which detect memory-safety bugs by instrumenting a program such that a memory violation crashes the process. Bug oracles are effective for a limited set of vulnerability classes, but they do not generalize to vulnerability classes that lack a reliable bug oracle. They also require expert review to validate exploitability, since the test harness may require high privileges and the bug may not be an exploitable vulnerability (for example, system-level mitigations may prevent exploitation, and some bugs are challenging to exploit on their own, such as an out-of-bounds read of a single byte).

We define a \textit{proof of exploitability} (PoE) as an end-to-end test that demonstrates an attacker capability against a deployed system, such as code execution, file read or write, authentication bypass, or denial of service. In this work, we create validation environments and exploitability oracles to verify such a capability without requiring a precise bug oracle. PoEs must demonstrate an attack in a realistic environment and may require a chain of multiple vulnerabilities.

\subsection{Known Exploited Vulnerabilities (KEVs)}

The Known Exploited Vulnerabilities (KEV) catalog~\cite{cisa_kev} is a curated database maintained by the U.S. Cybersecurity and Infrastructure Security Agency (CISA) containing flaws that are actively exploited in real-world attacks, including data breaches and ransomware incidents. In this work, we curate a benchmark of KEVs to demonstrate the effectiveness of our approach in finding vulnerabilities with real-world impact.

\section{Design}
\label{sec:design}

\sysname{} is designed to find exploitable vulnerabilities in large-scale systems with high recall, automatic validation (minimal manual review), and low inference and compute cost compared to direct LLM scanning. It consists of three subsystems: \textit{Synthesis}, \textit{Detection}, and \textit{Validation}, as shown in Figure~\ref{fig:architecture}. \textit{Synthesis} generates and refines static detectors once per vulnerability class, \textit{Detection} runs them on a target repository to find vulnerability candidates, and \textit{Validation} generates and checks a proof of vulnerability and a proof of exploitability for each candidate. Candidate detection and validation tasks are embarrassingly parallel, and \sysname{} runs them with concurrent workers.

\sysname{} scans generate the following three artifacts:
\begin{itemize}
\item \textbf{Static hits.} Synthesized detectors detect vulnerability patterns from source code analysis.
\item \textbf{Proofs of vulnerability.} For each static hit, \sysname{} generates an executable proof and accepts it only if the PoV environment's crash oracle observes the target's violation of the reported security property.
\item \textbf{Proofs of exploitability.} For each verified proof of vulnerability, \sysname{} generates an executable proof that demonstrates an attacker capability and the proof-of-exploitability oracle verifies it.
\end{itemize}

\subsection{Synthesis}
\label{sec:patterns}

The \textit{Synthesis} subsystem uses LLM agents to synthesize static detectors that identify vulnerability candidates in a target repository. Its design goal is to generate detectors that generalize across programming languages, frameworks, and vulnerability classes. To achieve this goal, \sysname{} uses a vulnerability dataset that we curate from public vulnerability reports and fixes. The dataset includes vulnerable versions and ground-truth labels that specify vulnerable source code locations for recall measurement. \sysname{} updates its detectors whenever a new vulnerability is added to the dataset.

To scan large systems effectively, \sysname{} refines the following three detectors:

\begin{enumerate}
    \item The \textit{Scope} detector decomposes a repository into subsystems. \sysname{} uses it to improve performance by building \ecpg{}s for each subsystem rather than building one for an entire system, as smaller \ecpg{}s are faster to analyze.
    \item The \textit{Attack Surface} detector locates entry points that cross a trust boundary, including HTTP, RPC, and framework callbacks. \sysname{} uses it to reduce the search space and focus on code paths that attackers can reach.
    \item The \textit{Candidates} detector searches for vulnerability candidates in paths that the attack surface reaches in each scoped \ecpg{}. Each detector uses \ecpg{} traversals to model a vulnerability class.
\end{enumerate}

Rather than optimizing for precision, \sysname{} optimizes for recall and reduces the resulting false positives with structural checks and iterative refinement. Each refinement round samples a detector's hits, classifies false positives that do not fit the structural requirements of the vulnerability class, and fixes the detector until it achieves high recall (zero false negatives) with low noise (few false positives). \sysname{} uses LLM agents and deterministic tests to verify that the detectors model a vulnerability class without overfitting to a specific CVE or project during synthesis.

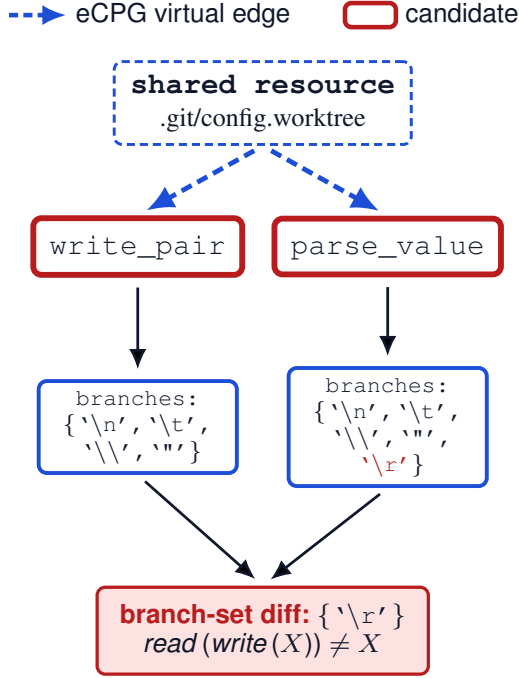
\begin{figure}[!t]
\centering
\resizebox{\columnwidth}{!}{\input{figures/cpg_traversal}}
\caption{An example of a detector that uses \ecpg{} traversals to find a semantic vulnerability, KEV \cve{2025}{48384} in Git. The root cause is an asymmetry between a configuration reader and a writer. \sysname{} adds a virtual node for the shared \code{.git/config} resource, links the writer (\code{write\_pair}) and reader (\code{parse\_value}) through it, and detects the asymmetry in their escape-character handling.}
\label{fig:cpg-traversal}
\end{figure}

\subsection{Detection}
\label{sec:detection}

The \textit{Detection} subsystem runs the synthesized detectors against the target repository. Detectors are heuristics for known vulnerability patterns that \sysname{} optimizes for high recall and low noise. The synthesized detectors use \ecpg{} traversals with detector-specific virtual edges and observations that over-approximate potential vulnerability candidates to achieve high recall.

We demonstrate an example of effective detection via \ecpg{} traversals of source-to-sink (taint-style) vulnerabilities. \textit{Source-to-sink} is a vulnerability pattern where attacker-controlled input (source) can reach a dangerous function call (sink). Figure~\ref{fig:motivation} shows our motivating example, a KEV in \code{telnetd} that static taint analysis tools fail to detect without a custom taint propagation model. \sysname{} overcomes this challenge by using breadth-first search (BFS) reachability instead of taint analysis: for every sink, it computes the set of methods that reach it, $R_S$, as well as the set of methods that can be reached from attacker-controlled entry points $R_E$, and adds the sink as a candidate if it is attacker-reachable, that is, $R_S \cap R_E \ne \emptyset$. For the motivating example, \code{net\_read} reaches a \code{setenv} that writes the \code{USER} environment variable, and a \code{getenv} that reads the same \code{USER} variable reaches the \code{execv} sink. The detector links this write and read with a virtual edge over the shared variable, and uses \ecpg{} traversals to detect the vulnerable flow. \sysname{} additionally detects dynamic dispatch and polymorphism patterns in dynamic languages such as Python and JavaScript by over-approximating with BFS over \ecpg{} virtual edges. As shown in Figure~\ref{fig:cpg-traversal}, \sysname{} uses \ecpg{} traversals with virtual nodes and edges to detect semantic flaws that taint analysis cannot model.

To achieve scalable continuous scanning, \sysname{} uses several optimizations. First, since \ecpg{} traversals use BFS rather than data-flow propagation, \sysname{} avoids constructing the Program Dependence Graph (PDG), which is typically more expensive to construct than the Abstract Syntax Tree (AST) or Control Flow Graph (CFG). Second, it extensively uses caching, for example to store the call graph, attack surface, and reachability trees. Third, it compiles the detectors and precomputes shared state. Finally, it uses a native engine rather than Joern, both to model languages that Joern does not support (such as Erlang) and to reduce scan time by an order of magnitude (Section~\ref{sec:efficiency}).

\begin{algorithm}[t]
\caption{\sysname{} end-to-end vulnerability discovery}
\label{alg:antiproof}
\footnotesize
\begin{algorithmic}[1]
\Statex \textbf{Input:} vulnerability dataset $\mathcal{D}$, target repository $R$, oracle library $\mathcal{O}$
\Statex \textbf{Output:} validated bugs $\mathcal{B}$, validated exploits $\mathcal{X}$
\Statex \textit{Synthesis (once, reused across repositories):}
\State $(\textsc{Scope}, \textsc{AttackSurface}, \textsc{Candidates}) \gets \Call{Synthesize}{\mathcal{D}}$
\Statex \textit{Discovery (per repository $R$):}
\State $(\mathcal{B}, \mathcal{X}) \gets (\emptyset, \emptyset)$
\State $S \gets \textsc{Scope}(R)$ \Comment{scoped subsystems}
\State $G \gets \Call{BuildCPG}{R, S}$ \Comment{scoped graph}
\State $A \gets \textsc{AttackSurface}(G)$ \Comment{trust-boundary entry points}
\State $\mathcal{C} \gets \textsc{Candidates}(G, A)$
\Statex \textit{Validation:}
\State $\mathcal{E} \gets \Call{PrepareValidationEnvs}{R, \mathcal{O}}$
\ForAll{$c \in \mathcal{C}$}
    \State $E_{\mathrm{PoV}} \gets \Call{PreparePoVEnv}{c}$
    \State $E \gets \Call{SelectValidationEnv}{\mathcal{E}, c}$
    \State $\pi_v \gets \Call{GeneratePoV}{E_{\mathrm{PoV}}, c}$
    \State $\pi_e \gets \Call{GeneratePoE}{E, c}$
    \If{$\Call{CheckProof}{E_{\mathrm{PoV}}, \pi_v} = 1$}
        \State $\mathcal{B} \gets \mathcal{B} \cup \{(c, \pi_v)\}$
    \EndIf
    \If{$\Call{CheckProof}{E, \pi_e} = 1$}
        \State $\mathcal{X} \gets \mathcal{X} \cup \{(c, \pi_e)\}$
    \EndIf
\EndFor
\State \Return $(\mathcal{B}, \mathcal{X})$
\end{algorithmic}
\end{algorithm}

\subsection{Validation}
\label{sec:poe}

The \textit{Validation} subsystem determines whether a vulnerability candidate is exploitable in a realistic target configuration. \sysname{} prepares a validation environment that is reused across candidates with the same target configuration and oracle. For each candidate, it also prepares an instrumented PoV environment with a crash oracle. \sysname{} generates PoVs and PoEs and checks each proof with its environment's oracle. These validation environments can also be used to automatically validate candidates produced by other vulnerability detection systems.

\textbf{Triage.} To manage the cost of LLM inference, \sysname{} supports LLM-based triage of static analysis findings to deduplicate, rank, and classify findings. This allows \sysname{} to prioritize realistic findings before investing in environment preparation or proof generation. The triage LLM agent can use the threat model of the system for classification, if available, to ignore findings outside of the threat model. The system can choose whether to remove findings, which may reduce recall, or to prioritize them within a budget.

\textbf{Validation Environment Preparation.} \sysname{} uses an LLM agent and reusable templates to prepare a reusable \textit{validation environment} per target configuration to validate vulnerability candidates. A validation environment includes an execution environment, the deployed system, and a proof-of-exploitability oracle. The execution environment may be a Docker container (e.g., for web applications), a virtual machine with GPUs (e.g., for LLM inference serving), or an emulator such as QEMU (e.g., for operating system kernels such as Linux). \sysname{} verifies that the deployed system version, dependencies, feature flags, and deployment options match the vulnerability candidate using deterministic verifiers. For each candidate, \sysname{} also prepares a PoV environment with a crash oracle and instruments the system to crash when it detects a violation of the relevant security property via a runtime monitor implemented using memory sanitizers (for memory-safety bugs) and instrumented assertions (for other vulnerability classes). The assertions allow the runtime monitor to detect semantic flaws by crashing the program when the security invariant is violated. \sysname{} refines the environments with feedback from an LLM verifier that checks consistency and realism.

\begin{figure}[t]
\centering
\resizebox{\columnwidth}{!}{\input{figures/oracle_challenge}}
\caption{An exploitability oracle as a challenge--response protocol. The oracle generates a challenge in the validation environment. The Prover, an LLM agent, connects over the network and executes a proof of exploitability against the target system. The oracle then verifies the state of the validation environment and accepts the proof if and only if the observed response correctly solves the challenge.}
\label{fig:oracle-challenge}
\end{figure}
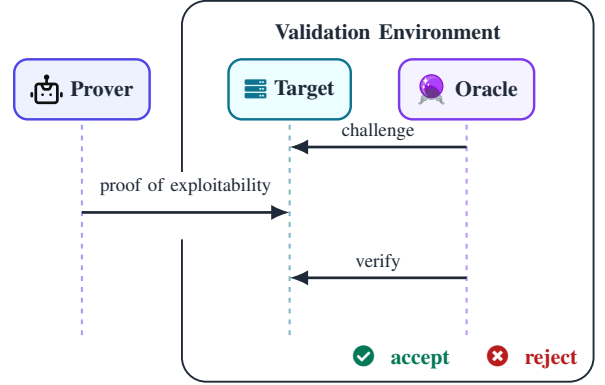

\textbf{Proof Generation.} As shown in Figure~\ref{fig:oracle-challenge}, the validation environment runs the target system alongside an exploitability oracle, a small trusted program that checks a single attacker capability, for example, code execution, arbitrary file access, SSRF, authentication bypass, or service crash. The Prover, an LLM agent, generates and executes a proof against the validation environment. The oracle independently checks the resulting environment state and accepts the proof only if it demonstrates the required attacker capability. This validation is inspired by CTF (Capture The Flag) challenge validation~\cite{shao2024nyu,zhu2025cve}, such as kernelCTF~\cite{kernelctf}, which verifies whether an attacker can achieve code execution on the Linux kernel. We extend the notion of CTF validation to capability-based challenges that prove attacker primitives beyond code execution and crashes. With a deterministic verifier, proof generation becomes a search problem, allowing LLMs or fuzzers to validate vulnerabilities beyond memory safety and capabilities beyond code execution.

Proof-of-exploitability validation shifts the review bottleneck from each vulnerability candidate to a smaller set of trusted oracles (once per capability) and reusable validation environments (once per target configuration). A single exploitability oracle can automatically validate many vulnerabilities and save manual review effort. A vendor that trusts a validation environment can use it to automatically validate vulnerabilities in their system, similar to how Google automatically validates Linux kernel vulnerability reports using kernelCTF (but not limited to code execution).

To ensure that proof-of-exploitability oracles accept a proof if and only if it demonstrates the attacker capability, we carefully review them using LLM agents and manual review. \sysname{} implements exploitability oracles that use a cryptographically secure secret $s$ and a mounted challenge directory \code{/challenge}, for example:

\begin{itemize}
    \item \textbf{Arbitrary Code Execution.} The oracle generates $s$, ensures that \code{/challenge/secret} does not exist, creates a script \code{/challenge/rce.sh} such that executing the script would write the secret $s$ into \code{/challenge/secret} (e.g., \code{echo "\$secret" > /challenge/secret}), then accepts a proof if and only if $s$ is indeed written to \code{/challenge/secret}.
    \item \textbf{Arbitrary File Read.} The oracle generates $s$ and stores it in \code{/challenge/secret}, then accepts a proof if and only if it returns $s$.
    \item \textbf{Arbitrary File Write.} The oracle generates $s$, ensures \code{/challenge/secret} does not exist, and provides it to the prover, then accepts a proof if and only if $s$ is indeed written to \code{/challenge/secret}.
    \item \textbf{Server-Side Request Forgery.} The oracle generates $s$ and provides it to the prover, runs a listener in the validation environment, and registers the listener's host under a list of aliases on the environment network. It accepts a proof if and only if the listener receives $s$ from the target rather than directly from the prover.
    \item \textbf{Authentication Bypass.} The oracle generates $s$ and places a proxy in
    front of the protected endpoints that returns $s$ when a request reaches a protected endpoint without authentication. The oracle accepts if and only if $s$ appears in the proof's output.
    \item \textbf{Denial of Service.} The oracle verifies a proof by checking whether the target has
    crashed (e.g., its process has died), become unavailable (e.g., its health check fails), or failed to terminate (and is killed after a timeout).
  \end{itemize}

Formally, let $E$ be a validation environment, let $C$ be an attacker capability (primitive), and let $\pi$ be a candidate executable proof. We define the oracle $O_C$ as:
\[
O_C(E, \pi) =
\begin{cases}
1 & \text{if $\pi$ demonstrates $C$ against $E$,}\\
0 & \text{otherwise.}
\end{cases}
\]
A rejected proof, $O_C(E, \pi) = 0$, does not imply the absence of $C$, only that the specific proof failed to demonstrate it. The exploitability oracles are independent of vulnerability classes by construction and depend only on the capability $C$ and the validation environment $E$. The oracles are sound if the security games accurately model the system and every accepted proof demonstrates the capability.

\textbf{Oracle threat model.} The exploitability oracle and validation environment are trusted and manually reviewed, while the prover is an untrusted attacker that connects to the target over the network. Each challenge uses a cryptographically secure secret and access-controlled challenge state. We manually review every validation environment to ensure that the prover cannot generate a proof that the oracle accepts without demonstrating the attacker capability, as guessing the secret has negligible success probability.

\textbf{Oracle coverage.} A single exploitability oracle can validate many different vulnerabilities. For example, the Arbitrary Code Execution oracle can validate 9 of the 2025 Top 10 KEV vulnerability classes, as shown in Table~\ref{tab:vuln-coverage}, including unsafe deserialization, OS command injection, and code injection vulnerabilities. It does not currently support client-side vulnerabilities such as cross-site scripting (XSS), since client-side code may be attacker-controlled. Because there are fewer exploitability oracles than vulnerability classes, manual review can focus on verifying the correctness of a small set of validation environments and automatically validate any vulnerability that is accepted by the exploitability oracles.

\begin{table}[!t]
\centering
\caption{\sysname{}'s coverage of the 2025 CWE Top 10 KEV Weaknesses~\cite{mitre_cwe_kev_top10_2025}.}
\label{tab:vuln-coverage}
\footnotesize
\setlength{\tabcolsep}{6pt}
\renewcommand{\arraystretch}{1.25}
\definecolor{kevok}{HTML}{2D6A4F}
\definecolor{kevno}{HTML}{B23A3A}
\begin{tabularx}{\columnwidth}{@{}>{\raggedright\arraybackslash}X c@{}}
\toprule
\textbf{Most actively exploited weakness} & \textbf{\sysname{}} \\
\midrule
OS command injection       & {\color{kevok}$\checkmark$} \\
Use after free             & {\color{kevok}$\checkmark$} \\
Out-of-bounds write        & {\color{kevok}$\checkmark$} \\
Missing authentication     & {\color{kevok}$\checkmark$} \\
Unsafe deserialization     & {\color{kevok}$\checkmark$} \\
Path traversal             & {\color{kevok}$\checkmark$} \\
Code injection             & {\color{kevok}$\checkmark$} \\
Authentication bypass      & {\color{kevok}$\checkmark$} \\
Heap-based buffer overflow & {\color{kevok}$\checkmark$} \\
Cross-site scripting       & {\color{kevno}$\times$} \\
\bottomrule
\end{tabularx}
\end{table}

\section{Implementation}
\label{sec:implementation}

We implemented our approach to support interchangeable static detectors and LLM agents. The static detectors output SARIF (Static Analysis Results Interchange Format) and the LLM agents output structured JSON. \sysname{} uses a PostgreSQL database and a web interface to store and deduplicate findings and track triage decisions, validation results, and disclosure status.

\textbf{Static Detectors.} \sysname{} synthesizes static detectors and refines them to be high-recall heuristics that approximate vulnerability candidates. It supports two \ecpg{} query engines: (1) Joern, which executes queries as Scala scripts using a Java Virtual Machine (JVM) per invocation, and (2) \enginename{}, our native Rust \ecpg{} engine implementation. \enginename{} partially implements the CPG specification~\cite{yamaguchi2014modeling, joerncpgspec} but does not construct the expensive Program Dependence Graph (PDG). It implements language frontends using \code{tree-sitter}~\cite{treesitter}, including languages that Joern does not officially support, such as Erlang. \sysname{} automatically detects subsystems, builds a CPG for each subsystem, and detects candidates using the selected \ecpg{} engine.

\textbf{Detector coverage.} \sysname{}'s \numdetectors{} synthesized detectors cover \numcoveredclasses{} vulnerability classes, including 9 of the 10 most actively exploited weakness classes, as shown in Table~\ref{tab:vuln-coverage}. The current implementation does not yet cover client-side classes such as cross-site scripting, which we leave to future work.

\textbf{Agents.} \sysname{} supports any LLM agent, including harnesses such as Claude Code and Codex. The system enforces structured outputs (JSON) using \code{pydantic} schemas and iterative refinement with verifier feedback. Agents are given a shell tool in a Docker container with language runtimes installed (e.g., Python 3.12, JDK 21, Erlang/OTP, Codex CLI 0.128.0, and Claude Code CLI 2.1.132 pre-installed), with \code{/code} mounted read-only and \code{/workspace} writable and reset between candidates.

\textbf{Validation Environments.} \sysname{} implements validation environments using Docker containers, QEMU (e.g., for operating system emulation), or virtual machines (e.g., GPUs for inference serving). Each exploitability oracle generates a challenge for one attacker capability and checks whether the LLM-generated proof solves it. \sysname{} creates a container for generated proofs and connects it to the validation environment through a Docker network or SSH forwarding. For PoVs, \sysname{} uses the crash oracle and creates a candidate-specific PoV environment with assertions that crash when the vulnerability is detected.

\textbf{Vulnerability Dataset.} We implemented scripts that collect vulnerability reports from GitHub Security Advisories, bug bounty writeups, and fix commits. We collected over 400 CVE reports in the evaluated projects to analyze common vulnerability patterns. For recall measurement, we collect the reports of the benchmark vulnerabilities and generate ground-truth labels for static detection that score a hit based on line proximity or enclosing method.

\section{Evaluation}
\label{sec:evaluation}

We evaluate \sysname{} using the following research questions:

\begin{itemize}
    \item \textbf{RQ1.} What recall does \sysname{} achieve on known exploitable vulnerabilities? (Section~\ref{sec:benchmarks})
    \item \textbf{RQ2.} Can \sysname{} discover previously unknown vulnerabilities? (Section~\ref{sec:zeroday})
    \item \textbf{RQ3.} How efficient is \sysname{} at scanning large systems? (Section~\ref{sec:efficiency})
    \item \textbf{RQ4.} Which components contribute to recall and cost? (Section~\ref{sec:ablations})
\end{itemize}

\subsection{Experimental Setup}
\label{sec:eval-setup}

All experiments run on a GCP \code{n2-highmem-32} instance (32 vCPUs, 256\,GB RAM) provisioned via SkyPilot~\cite{yang2023skypilot}. Experiments that include end-to-end validation that requires GPUs, for example inference engines, additionally use GCP \code{g2-standard-4} instances with one NVIDIA L4 GPU.

\textbf{Baselines.} For RQ1, our primary comparisons are open-source static-analysis and neuro-symbolic tools. We run the baselines in Docker containers that mount the target at \code{/code} and write tool outputs to \code{/workspace}. We score each baseline by the number of findings that are semantically equivalent to the benchmark ground-truth vulnerability reports. We compare recall against the following baselines:
\begin{itemize}
    \item \textbf{Semgrep}~\cite{semgrep_docs}. Semgrep is an open-source static analysis tool that scans code for security vulnerabilities using predefined rules. We use the open-source \code{semgrep/semgrep} Docker image and run \code{semgrep scan} with \code{--config=auto}, \code{--no-git-ignore}, and JSON output.
    \item \textbf{CodeQL}~\cite{avgustinov2016ql}. CodeQL is a semantic code analysis engine that models vulnerabilities as queries over program structure and data flow. We use GitHub CodeQL CLI 2.25.5, create databases with \code{autobuild}, and run the language-specific \code{security-extended} query suite.
    \item \textbf{RepoAudit}~\cite{guo2025repoaudit}. RepoAudit is an LLM-guided repository-auditing system for source-to-sink bug patterns. We run its \code{dfbscan} mode with reachability enabled, temperature 0, and call depth 3. RepoAudit only supports three vulnerability classes (null-pointer dereference, memory leak, and use-after-free).
    \item \textbf{KNighter}~\cite{yang2025knighter}. KNighter synthesizes static-analysis checkers from vulnerability-fixing commits. KNighter only supports Linux-kernel vulnerabilities and struggles with use-after-free vulnerabilities that require state-machine and concurrency reasoning. In our benchmarks, this leaves only one supported target, the Linux use-after-free KEV. We provide KNighter with similar vulnerability patches and measure whether the generated detector can find the Linux KEV.
\end{itemize}

We also include direct LLM scanning reference points using Codex with \code{gpt-5.3-codex}, \code{gpt-5.4-mini}, and \code{gpt-5.5}, and Claude Code with \code{claude-opus-4.6}. We include \code{claude-opus-4.6} because Anthropic uses it in their published large-scale zero-day discovery evaluation in open-source software~\cite{anthropic_zero_days_2026,anthropic_firefox_2026}. These agents are not primary baselines because their performance depends strongly on model choice, prompt engineering, and inference budget. We run each agent in a Docker container with its CLI, Python 3.12, \code{git}, \code{ripgrep}, \code{jq}, and \code{uv}. The target repository is mounted read-only at \code{/code} with a writable \code{/workspace}. The read-only mount prevents the agent from modifying the project's code, and the workspace allows the agent to save proof-of-concept artifacts and temporary files. As shown in Appendix~\ref{appendix:prompts}, we use prompts that instruct the agent to find all vulnerabilities in a candidate file and to report each finding with a proof-of-concept reproduction, similar to the approach described by Anthropic for their zero-day scans~\cite{anthropic_zero_days_2026}. We do not evaluate Claude Code Security or Codex Security because their terms of service do not allow scanning third-party repositories.

We allow 30 minutes per target and run each configuration 10 times. Table~\ref{tab:baselines} shows the mean detections rounded to the nearest vulnerability and total recall percentages computed from the unrounded means.

\subsection{Metrics}

We use the following metrics in our evaluation:

\textbf{Recall.} For known vulnerabilities, we measure \textit{recall} as the fraction of benchmark vulnerabilities for which the system detects the ground-truth vulnerability.

\textbf{Hits, PoVs, and PoEs.} For zero-day discovery, we track the funnel of \emph{Hits} (static-detector candidates), \emph{PoVs} (proofs of vulnerability that the crash oracle accepts in a candidate-specific PoV environment), and \emph{PoEs} (proofs of exploitability accepted by the exploitability oracle).

\textbf{Cost.} We measure the time and cost in USD of candidate discovery (CPG construction and static detector execution) and automatic validation (environment preparation, environment execution, and LLM inference).

\begin{table}[!t]
\centering
\caption{Recall on BountyBench and KEVBench.}
\label{tab:baselines}
\scriptsize
\setlength{\tabcolsep}{2.0pt}
\renewcommand{\arraystretch}{1.12}
\begin{tabular}{@{}llccc@{}}
\toprule
\textbf{Method} & \textbf{Model} & \shortstack{\textbf{BountyBench}\\$(n=\bbsupported{})$} & \shortstack{\textbf{KEVBench}\\$(n=\kevtotal{})$} & \shortstack{\textbf{Total}\\$(n=66)$} \\
\midrule
Semgrep & -- & 12 & 5 & 17 (26\%) \\
CodeQL & -- & 6 & 0 & 6 (9\%) \\
RepoAudit & \code{gpt-5.5} & 0 & 0 & 0 (0\%) \\
KNighter & \code{gpt-5.5} & 0 & 0 & 0 (0\%) \\
\midrule
Claude Code & \code{claude-opus-4.6} & 31 & 8 & 39 (59\%) \\
Codex & \code{gpt-5.3-codex} & 27 & 12 & 39 (58\%) \\
Codex & \code{gpt-5.4-mini} & 27 & 11 & 38 (58\%) \\
Codex & \code{gpt-5.5} & 27 & 10 & 37 (56\%) \\
\midrule
\sysname{} & \code{claude-opus-4.6} & 42 & 19 & 61 (92\%) \\
\sysname{} & \code{Qwen 3.7-Max} & 41 & 19 & 60 (91\%) \\
\sysname{} & \code{GLM-5.1} & 42 & 17 & 59 (89\%) \\
\sysname{} & \code{gpt-5.3-codex} & 42 & \textbf{20} & 62 (94\%) \\
\textbf{\sysname{}} & \textbf{\code{gpt-5.5}} & \textbf{44} & \textbf{20} & \textbf{64 (97\%)} \\
\bottomrule
\end{tabular}
\renewcommand{\arraystretch}{1.0}
\end{table}

\subsection{RQ1: Known-Vulnerability Detection}
\label{sec:benchmarks}

We evaluate \sysname{} on known-vulnerability benchmarks to measure recall. First, we evaluate on BountyBench, a public benchmark of real-world bug-bounty vulnerabilities. To reduce data-contamination risk and focus on exploitable vulnerabilities with real-world impact, we additionally curate KEVBench, a dataset of CISA Known Exploited Vulnerabilities~\cite{cisa_kev} that became public in 2025 and 2026, after the knowledge cutoff of some of the LLMs we use in our evaluation.

\textbf{Benchmark scope.} We measure recall on vulnerability detection, where the system generates vulnerability reports given a project. We do not evaluate on CyberGym~\cite{wang2025cybergym} because its primary task is proof-of-concept reproduction from a vulnerability description, and its benchmark is centered on memory-safety bugs in fuzzing-style targets, while we focus on broad vulnerability-class coverage beyond memory safety.

\textbf{BountyBench.} BountyBench~\cite{zhang2026bountybench} contains real-world vulnerabilities from 25 open-source projects, curated from public bug bounty programs with disclosed awards totaling over USD~60{,}000. To the best of our knowledge, the highest previously reported detection rate on BountyBench is Co-RedTeam's \bbsota\% result with \code{Gemini-3-pro}~\cite{he2026co}. We use BountyBench as a known-vulnerability target set by collecting its CVEs, checking out the affected project versions, and scoring whether each system reports a finding semantically equivalent to the ground-truth vulnerability. \sysname{} detects \bbdetected{} of \bballtotal{} BountyBench vulnerabilities (\bbdetectrate\%).

\textbf{KEVBench.} We curate KEVBench from CISA's Known Exploited Vulnerabilities~\cite{cisa_kev} (KEVs) catalog by selecting 2025 and 2026 KEVs that affect open-source projects. The resulting benchmark contains \kevtotal{} exploited-in-the-wild vulnerabilities across 12 vulnerability classes, 7 programming languages, and 17 projects.

\textbf{Results.} As shown in Table~\ref{tab:baselines}, \sysname{} detects \numdetected{} of \numtargets{} vulnerabilities with \code{gpt-5.5}, outperforming the strongest primary baseline, Semgrep, by 71 percentage points and the strongest direct LLM-agent reference point by 38 percentage points. It detects 62, 61, 60, or 59 of \numtargets{} using \code{gpt-5.3-codex}, \code{claude-opus-4.6}, \code{Qwen 3.7-Max}, or \code{GLM-5.1}, respectively. Because direct LLM scanning depends on the model, prompt, tool setup, and inference budget, we do not claim that LLM agents cannot find these vulnerabilities. \sysname{} is complementary to direct LLM scanning because its static detectors can guide LLM agents during search, and its exploitability oracles can validate findings produced by those agents.

\subsection{RQ2: Zero-Day Discovery}
\label{sec:zeroday}

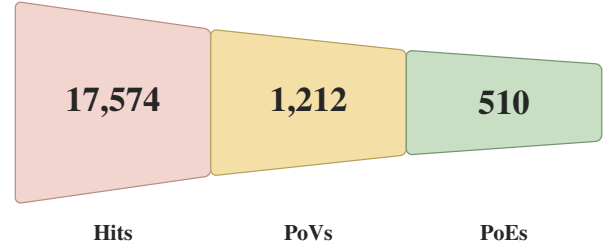
\begin{figure}[!t]
\centering
\resizebox{\linewidth}{!}{\input{figures/zeroday_funnel}}
\caption{End-to-end zero-day discovery funnel across \numprojects{} scanned projects.}
\label{fig:zeroday-funnel}
\end{figure}

We evaluate \sysname{} on \numprojects{} widely deployed open-source projects. We select projects with high security impact on AI and cloud infrastructure, including LLM training and inference systems, that can be emulated in Docker containers, QEMU, or virtual machines. The selected projects total 136 million lines of code across \numlanguages{} languages, with an average of 2.7 million lines per project and a maximum of 31.5 million lines.

\textbf{Results.} In our initial zero-day scan, \sysname{} found \numcandidates{} vulnerability candidates, generated \numfindings{} PoVs, and accepted \numpoes{} PoEs (Figure~\ref{fig:zeroday-funnel}). Due to time constraints, we selected \nummanualverified{} findings for manual review and coordinated disclosure. \sysname{} has uncovered critical and high-severity vulnerabilities, including remote code execution in LLM training and inference systems, and denial-of-service and potential code-execution flaws in critical infrastructure such as OpenSSL and FreeBSD. We have received \numcves{} CVE assignments to date and confirmed \numduplicates{} vulnerabilities before their CVEs were publicly disclosed. We continue to improve the static detectors in recall and precision and to scale the validation environments. Table~\ref{tab:scan-stats} reports the per-project language, size, funnel, and cost.

\textbf{Cost reference.} Anthropic reports that the OpenBSD campaign that found the SACK vulnerability ran Mythos roughly a thousand times at a total cost of USD~20{,}000~\cite{anthropic_mythos_preview_2026}. \sysname{} validated its static findings in OpenBSD with \code{gpt-5.3-codex} for less than USD~612, a $33\times$ lower cost, and reproduces the Mythos findings, SACK null-dereference in OpenBSD and FreeBSD \cve{2026}{4747}, for USD~0.46 and USD~1.73, respectively. These costs reflect a single run of an early prototype and represent an upper bound, as we have continued to optimize the system since the scan.

\begin{table}[!t]
\centering
\caption{\sysname{}'s zero-day scan across \numprojects{} widely deployed projects.}
\label{tab:scan-stats}
\scriptsize
\setlength{\tabcolsep}{3pt}
\renewcommand{\arraystretch}{1.0}
\begin{tabular}{@{}llrrrrr@{}}
\toprule
\textbf{Project} & \textbf{Language} & \textbf{LOC} & \textbf{Hits} & \textbf{PoV} & \textbf{PoE} & \textbf{Cost (USD)} \\
\midrule
openbsd & C & 16.1M & 4{,}118 & 190 & 65 & 611.90 \\
freebsd & C & 19.5M & 3{,}828 & 180 & 3 & 460.33 \\
fluent\_bit & C & 806k & 1{,}208 & 87 & 1 & 315.06 \\
mariadb & C++ & 1.9M & 1{,}069 & 100 & 63 & 140.98 \\
postgres & C & 1.8M & 782 & 25 & 6 & 69.34 \\
mlflow & Python & 1.6M & 554 & 51 & 30 & 74.06 \\
transformers & Python & 1.3M & 514 & 26 & 6 & 18.42 \\
openssl & C & 993k & 513 & 12 & 1 & 84.80 \\
sglang & Python & 1.1M & 472 & 62 & 35 & 75.37 \\
linux & C & 31.5M & 466 & 33 & 6 & 76.33 \\
qemu & C & 2.4M & 460 & 36 & 0 & 148.01 \\
git & C & 1.1M & 440 & 10 & 3 & 61.82 \\
airflow & Python & 1.4M & 405 & 33 & 22 & 34.05 \\
vllm & Python & 1.0M & 251 & 56 & 49 & 61.90 \\
litellm & Python & 1.5M & 230 & 21 & 8 & 38.94 \\
httpd & C & 447k & 208 & 15 & 0 & 99.45 \\
moby & Go & 2.3M & 193 & 4 & 3 & 12.78 \\
openssh & C & 149k & 164 & 11 & 9 & 56.26 \\
redis & C & 386k & 164 & 4 & 2 & 16.19 \\
bentoml & Python & 92k & 146 & 38 & 28 & 22.02 \\
huggingface\_hub & Python & 91k & 140 & 29 & 14 & 23.57 \\
ollama & Go & 749k & 138 & 3 & 2 & 18.28 \\
argo\_cd & Go & 2.3M & 124 & 10 & 10 & 14.02 \\
llama\_cpp & C++ & 655k & 123 & 34 & 25 & 32.46 \\
tomcat & Java & 440k & 98 & 34 & 32 & 31.79 \\
containerd & Go & 1.2M & 83 & 14 & 13 & 13.82 \\
jenkins & Java & 289k & 81 & 2 & 1 & 10.73 \\
nccl & C++ & 132k & 72 & 23 & 8 & 83.32 \\
onnxruntime & C++ & 3.9M & 70 & 4 & 4 & 6.33 \\
nginx & C & 213k & 62 & 2 & 2 & 39.14 \\
kserve & Go & 5.4M & 50 & 22 & 22 & 10.01 \\
jupyterhub & Python & 66k & 48 & 4 & 4 & 7.34 \\
langfuse & TypeScript & 533k & 47 & 1 & 1 & 2.80 \\
gvisor & Go & 578k & 39 & 9 & 9 & 7.65 \\
buildkit & Go & 1.2M & 37 & 10 & 8 & 5.88 \\
harbor & Go & 341k & 35 & 2 & 0 & 2.72 \\
runc & Go & 324k & 33 & 0 & 0 & 5.57 \\
pytorch & Python & 3.2M & 31 & 6 & 6 & 20.23 \\
gitlab & Ruby & 8.2M & 29 & 5 & 5 & 7.03 \\
triton\_inference\_server & Python & 134k & 24 & 2 & 2 & 4.03 \\
ingress\_nginx & Go & 147k & 6 & 0 & 0 & 0.92 \\
nvidia\_container\_toolkit & Go & 890k & 6 & 0 & 0 & 0.88 \\
prometheus & Go & 447k & 5 & 0 & 0 & 0.56 \\
ray & Python & 1.4M & 4 & 2 & 2 & 0.72 \\
grafana & Go & 3.5M & 2 & 0 & 0 & 0.20 \\
safetensors & Rust & 9k & 2 & 0 & 0 & 0.00 \\
envoy & C++ & 1.5M & 0 & 0 & 0 & 0.00 \\
firecracker & Rust & 145k & 0 & 0 & 0 & 0.00 \\
kubernetes & Go & 5.4M & 0 & 0 & 0 & 0.00 \\
tensorflow & C++ & 5.1M & 0 & 0 & 0 & 0.00 \\
\midrule
\textbf{Total (50 projects)} & & \textbf{136M} & \textbf{\numcandidates{}} & \textbf{\numfindings{}} & \textbf{\numpoes{}} & \textbf{\numscancost{}} \\
\bottomrule
\end{tabular}
\end{table}

\subsection{RQ3: Detector Efficiency and Scalability}
\label{sec:efficiency}

\begin{figure}[!t]
\centering
\input{figures/detector_eval}
\caption{Detector efficiency and scalability on KEVBench and BountyBench.}
\label{fig:detector-eval}
\end{figure}
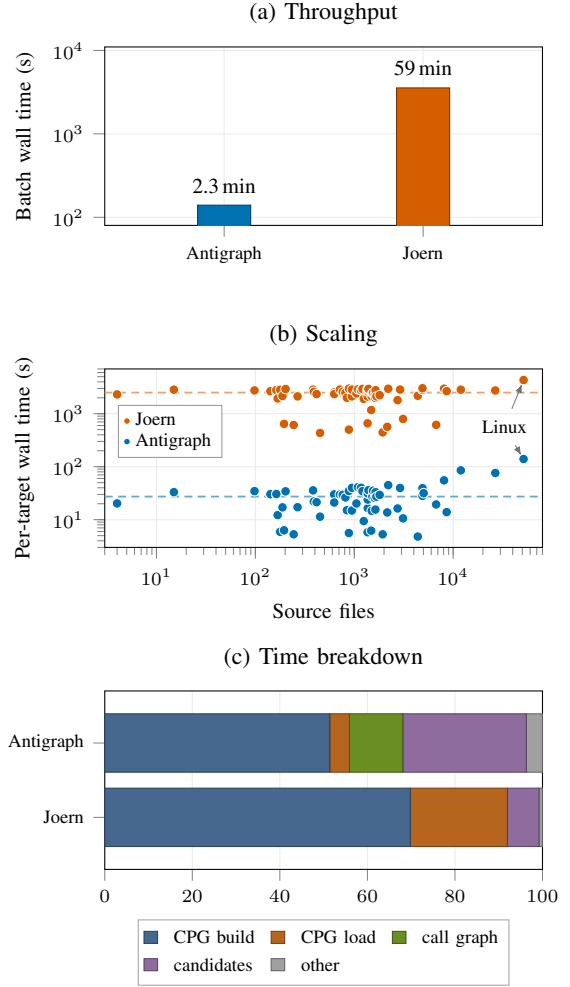

Continuous scanning of large systems requires fast detection that scales with project size. We compare \enginename{}, \sysname{}'s native \ecpg{} engine, against Joern on the full 66-target KEV and BountyBench set. Both engines run with a cold cache, rebuilding all graphs in every iteration, and reach identical recall (\numdetected{} of \numtargets{}).

\textbf{Throughput.} As Figure~\ref{fig:detector-eval}(a) shows, \enginename{} scans all benchmark targets \detspeedup$\times$ faster than Joern, a median of \detbatchmin{} versus \joernbatchmin{} minutes across 5 trials.

\textbf{Scaling.} As shown in Figure~\ref{fig:detector-eval}(b), \enginename{}'s median per-target time is 27.3\,s, 92$\times$ faster than Joern's 2{,}505\,s, with a maximum of 140\,s on the largest target, the Linux kernel at 51k scoped files.

\textbf{Time breakdown.} As shown in Figure~\ref{fig:detector-eval}(c), Joern spends 92\% of its run time building and loading CPGs. By contrast, \enginename{} is \detspeedup$\times$ faster, as it avoids building the expensive program-dependence graph via BFS reachability in Rust over the \ecpg{} rather than Joern's data-flow propagation in Scala.

\subsection{RQ4: Component Contributions}
\label{sec:ablations}

We run the following ablations to measure how \sysname{}'s components contribute to recall and cost:
\begin{itemize}
    \item \textbf{$-$\ecpg{}} replaces \sysname{}'s reachability traversal with Joern's native data-flow reachability, \code{reachableByFlows}.
    \item \textbf{$-$\ecpg{} frontends} disables the custom \ecpg{} frontends that \sysname{} implements for languages Joern cannot ingest natively, namely Erlang and Velocity.
    \item \textbf{$-$Attack Surface} disables the detector that selects attacker-facing entry points.
    \item \textbf{$-$Scope} disables repository-scope detection and builds whole-repository CPGs.
    \item \textbf{$-$observations} removes technique facts and structural observations from the prover prompt.
\end{itemize}
Static detector runs use a 20-minute per-target timeout.

\begin{table}[!t]
\centering
\caption{Component ablation study.}
\label{tab:ablations}
\scriptsize
\setlength{\tabcolsep}{2pt}
\renewcommand{\arraystretch}{1.14}
\resizebox{\columnwidth}{!}{%
\begin{tabular}{@{}lcccrr@{}}
\toprule
\textbf{Variant} & \shortstack{\textbf{BountyBench}\\$(n=46)$} & \shortstack{\textbf{KEVBench}\\$(n=20)$} & \shortstack{\textbf{Total}\\$(n=66)$} & \textbf{Candidates} & \textbf{Time} \\
\midrule
\multicolumn{6}{@{}l}{\textit{Detector ablations}} \\
\sysname{}                              & 44 & 20 & \textbf{64} & 32{,}705 & 140s \\
$-$\ecpg{}                    & 30 & 16 & 46          & 2{,}028 & 3{,}490s \\
$-$\ecpg{} frontends                  & 44 & 18 & 62          & 32{,}682 & 264s \\
$-$Attack Surface                       & 44 & 20 & 64          & 47{,}280 & 294s \\
$-$Scope                                & 44 & 20 & 64          & 160{,}005 & 456s \\
\midrule
\multicolumn{6}{@{}l}{\textit{Validation ablations}} \\
\shortstack[l]{$-$observations\\(\code{gpt-5.3-codex}, medium)} & 36 & 15 & 51 & -- & -- \\
\shortstack[l]{$-$observations\\(\code{gpt-5.5}, medium)} & 41 & 19 & 60 & -- & -- \\
\bottomrule
\end{tabular}%
}
\renewcommand{\arraystretch}{1.0}
\end{table}

\textbf{Results.} The \textit{Detection ablations} in Table~\ref{tab:ablations} show that each component improves recall or efficiency. \sysname{} with \ecpg{} detects 18 more benchmark vulnerabilities than Joern's native data-flow reachability and runs 25$\times$ faster. The custom \ecpg{} frontends for Erlang and Velocity recover two KEVBench vulnerabilities that Joern's native frontends cannot parse. Scope detection reduces the candidate count by 80\% and scan time by 3.3$\times$ by focusing CPG construction on relevant code, while maintaining recall. Attack Surface detection reduces the candidate count by 31\% and scan time by 2.1$\times$ while maintaining recall.

The \textit{Validation ablations} show that the static observations improve recall. Removing observations with \code{gpt-5.3-codex} at medium effort detects 36 of 46 BountyBench targets and 15 of 20 KEVBench targets, for 51 of \numtargets{} total. With \code{gpt-5.5} at medium effort, the system detects 41 of 46 BountyBench targets and 19 of 20 KEVBench targets, for 60 of \numtargets{} total.

\section{Discussion and Limitations}
\label{sec:discussion}

\textbf{Vulnerability Detectors.} \sysname{} can only find vulnerabilities that the synthesized detectors detect. Because the system synthesizes detectors from known vulnerability patterns, it may miss a novel pattern until it learns a corresponding detector. To mitigate such false negatives, \sysname{} can synthesize new detectors from additional known vulnerabilities or generate candidates with external systems, such as LLM agents or other static analysis tools. Moreover, \sysname{} uses synthesized detectors that detect scope and attack surface to avoid path explosion, which may lead to false negatives if a real vulnerability is outside of the detected paths. \sysname{} mitigates this by testing its detectors on a wide range of projects across different programming languages and frameworks.

\textbf{Static Analysis Tools.} \sysname{} depends on the quality of its static detectors and \ecpg{} engines. It supports two engines, Joern and \enginename{}, which must correctly parse the source languages, build the \ecpg{}, and execute its traversals. Any error in these steps can cause \sysname{} to miss vulnerabilities.

\textbf{End-to-end validation.} Automatic vulnerability validation requires environments that accurately simulate the target system. PoV environments are candidate-specific, whereas a validation environment can be reviewed once and reused across candidates. Our evaluation focuses on open-source projects because \sysname{} assumes source-code access for analysis and validation. We leave scanning proprietary systems to future work. In some cases, accurate simulation may be challenging, for example, when hardware emulation or third-party authentication is required, which may lead to false positives or false negatives. Our proof-of-exploitability oracles cover the attacker primitives evaluated in this paper, including remote code execution, arbitrary file access, SSRF, authentication bypass, and denial of service. Vulnerabilities whose impact does not map cleanly to these primitives may require experts to create and manually review new exploitability oracles.

\textbf{Threat Model.} \sysname{} may generate reports for exploitable vulnerabilities that are outside the threat model of the target system. To mitigate that, we use LLMs to generate and update a threat model document. \sysname{} uses the threat model document to classify and prioritize findings. Future work may improve the accuracy of these LLM-generated threat models, allowing \sysname{} to focus on in-scope findings and reduce cost and false positives. In some cases, project maintainers may dispute real vulnerabilities and consider them out of scope or without security impact, while real-world attackers exploit them in the wild~\cite{oligo2024shadowray}.

\textbf{LLM limitations.} \sysname{} uses LLMs to synthesize detectors, generate proof-of-vulnerability (PoV) tests, set up validation environments, and generate proofs of exploitability (PoE). Generating a PoE may be challenging or require a chain of multiple vulnerabilities. The model may also refuse cybersecurity tasks such as exploitation due to safety filters. To reduce these false negatives, \sysname{} supports any LLM agent, including open-weight models, and can run multiple agents with different providers. To reduce false positives, we manually review the exploitability oracles and validation environments.

\textbf{Threats to validity.} As our known-vulnerability benchmarks, BountyBench and KEVBench, are publicly available, LLM agents may memorize the vulnerabilities or exploits from their training data or discover them via web search. To mitigate this risk of data contamination, KEVBench uses known exploited vulnerabilities that became public in 2025 and 2026, after the knowledge cutoff of some of the LLMs we evaluate. To reduce the risk of memorization, we additionally evaluate \sysname{}'s effectiveness in zero-day discovery. Exact reproduction of our experiments that use proprietary LLM models may be challenging, as their performance can depend on the model capabilities and safety filters. To reduce this risk, \sysname{} supports open-weight models, such as \code{GLM-5.1}.

\section{Related Work}
\label{sec:related}

\textbf{Program Analysis.} Program analysis approaches for vulnerability discovery, such as static analysis~\cite{ayewah2008static,bessey2010few}, fuzzing~\cite{li2018fuzzing,manes2019art}, taint analysis~\cite{newsome2005dynamic}, and symbolic execution~\cite{cha2012mayhem,baldoni2018survey}, have been extensively studied. Query-based approaches, such as CodeQL~\cite{avgustinov2016ql} and code property graphs~\cite{yamaguchi2014modeling}, are effective for specific vulnerability patterns, but rely on expert-written models, source and sink specifications, and manual validation of alerts. Fuzzing and symbolic execution are effective for vulnerability discovery and automatic exploit generation~\cite{avgerinos2011aeg}, but are primarily limited to memory safety and struggle with harness engineering, exponential search spaces, and deep program states. Arbiter~\cite{vadayath2022arbiter} and SAILOR~\cite{shafiuzzaman2026sailor} combine static analysis with symbolic execution to discover specific memory safety vulnerabilities. In contrast, \sysname{} detects vulnerabilities beyond memory-safety and combines neuro-symbolic static analysis for high-recall detection with proofs of exploitability for automatic validation.

\textbf{LLM-Based Static Analysis.} Prior work uses LLMs to guide vulnerability classification with code property graphs (LLMxCPG~\cite{lekssays2025llmxcpg}) and to infer specifications for taint-based vulnerabilities in Java (IRIS~\cite{li2025iris}, Argus~\cite{liang2026argus}), PHP (Artemis~\cite{ji2025artemis}), and JavaScript (SemTaint~\cite{ghebremichael2026multi}). Recent work explores neuro-symbolic static analysis, where LLMs synthesize static detectors from historical vulnerability patterns, including C/C++ static analyzers (KNighter~\cite{yang2025knighter}), CodeQL queries (QLCoder~\cite{wang2025qlcoder}), and Joern queries (MoCQ~\cite{li2025automated}). ProtocolGuard~\cite{song2026protocolguard} combines LLM-guided static analysis with dynamic verification to detect protocol non-compliance bugs.
 In comparison, \sysname{} introduces \ecpg{} traversals to detect vulnerabilities that prior approaches fail to model, including semantic vulnerabilities, synthesizes static detectors and refines them for high recall on a wide range of vulnerability classes, and uses proof-of-exploitability oracles for automatic validation rather than expert review or LLM verifiers.

\textbf{Agentic Vulnerability Discovery.}
LLM agents have shown promise for vulnerability discovery, exploitation, and patching~\cite{rohlf2025lifecycle,zhang2025llms,sheng2025llms,zhou2025large,li2025everything,ding2024vulnerability,zhu2026teams,deng2024pentestgpt,singer2025incalmo,he2026co,du2024vul,wang2025vulagent,guo2025repoaudit,wang2025cybergym,shicybergym,darpaAixccWebsite,darpa2025aixccresults,heelan2026industrialisation}. Proprietary systems including Claude Mythos~\cite{anthropic_mythos_preview_2026,anthropic_glasswing_2026}, Claude Code Security~\cite{anthropic_zero_days_2026,anthropic_firefox_2026}, Codex Security~\cite{openai_codex_security_2026}, BigSleep~\cite{google2025summersecurity}, XintCode~\cite{xint2025xintcode}, and XBOW~\cite{xbow2025chaosphase} have used LLMs to uncover thousands of vulnerabilities in production systems. In contrast to direct LLM scanning, which is prohibitively expensive for continuous scanning and suffers from high false-positive rates~\cite{li2026llm}, \sysname{} uses efficient static detectors to continuously scan large systems and proofs of exploitability for automatic validation.

\textbf{Vulnerability Validation.} Proof-of-Vulnerability (PoV) test generation has been extensively studied~\cite{sheng2025llms}. Recent works, including AnyPoC~\cite{zhao2026anypoc}, FaultLine~\cite{nitin2025faultline}, and CVE-GENIE~\cite{ullah2025cve}, leverage LLMs for proof-of-vulnerability test generation. Because these approaches automatically validate the PoV using LLM-as-a-judge verifiers or bug oracles such as memory sanitizers~\cite{serebryany2012addresssanitizer,stepanov2015memorysanitizer,song2019sok}, they either are limited in vulnerability-class coverage or inherit LLM limitations such as hallucinations, reward hacking, and weaker performance on
long-context reasoning~\cite{kalai2025language,li2026llm}. Recent exploitability benchmarks such as ExploitBench~\cite{lee2026exploitbench} and ExploitGym~\cite{wang2026exploitgym} evaluate the ability of LLMs to exploit vulnerabilities, but focus on code execution from memory-safety vulnerabilities. Rather than relying on manual review or LLMs, \sysname{} extends the notion of CTF-based validation~\cite{kernelctf} to prove attacker capabilities without a precise bug oracle. This shifts the manual review bottleneck from every vulnerability candidate to a small set of validation environments and oracles and allows \sysname{} to automatically validate vulnerability classes beyond memory safety.

\section{Conclusion}
\label{sec:conclusion}

Scalable vulnerability discovery is critical to secure computer systems before attackers exploit them. In this paper, we have presented \sysname{}, an end-to-end vulnerability discovery system for continuously scanning large-scale projects for vulnerability patterns with high recall and low discovery and validation costs. \sysname{} uses LLMs to continually synthesize and refine static detectors for high-recall discovery and executable proofs of exploitability for automatic validation. We show that this novel approach outperforms existing baselines on two known vulnerability benchmarks, BountyBench and our curated KEVBench dataset. Evaluated on \numprojects{} widely deployed open-source projects, \sysname{} generated \numpoes{} proofs of exploitability. Due to time constraints, we selected \nummanualverified{} findings for manual review and coordinated disclosure. To date, we have received \numcves{} CVE assignments for vulnerabilities of critical or high severity, including remote code execution vulnerabilities that could lead to complete takeover of widely deployed LLM inference and training systems.

\section*{Ethics Considerations}

We recognize the dual-use risk of vulnerability discovery. To mitigate this risk, we manually review selected findings and coordinate disclosure of confirmed vulnerabilities with affected maintainers and vendors. When possible, we suggest patches or mitigations and allow affected parties sufficient time to develop and deploy fixes before public disclosure. We evaluate only on publicly available open-source projects and vulnerability benchmarks, and reproduce vulnerabilities only in local environments without affecting production systems. We do not collect personally identifiable information or sensitive data in this research. Our goal is to help defenders find and fix existing flaws before attackers exploit them.

\ifpublicversion
\section*{Acknowledgments}
We thank the students in the Sky Computing Lab security group for their feedback. This work is supported by gifts from Accenture, Algorithmic SuperIntelligence Labs, Amazon, AMD, Anyscale, Broadcom, cmpnd, Google, IBM, Intel, Intesa Sanpaolo, Lambda, Lightspeed, Mirendil, NVIDIA, Samsung SDS, and VESSL.
\fi

\bibliographystyle{IEEEtran}
\bibliography{references}

\appendices

\section{Agent Prompts}
\label{appendix:prompts}

We show simplified excerpts from the system prompts for \sysname{}'s LLM agents (\textit{Synthesis}, \textit{Environment Preparation}, \textit{Environment Verifier}, and \textit{Prover}) and the prompt used for the direct LLM-agent reference points. Bracketed ellipses mark omitted operational instructions.

\noindent\textbf{Synthesis agent.} Refines the detectors for one vulnerability class against the curated vulnerability dataset.

\begin{lstlisting}[style=prompt]
Your goal is to refine one vulnerability class's detector, written as an **eCPG traversal**: a structural query over the extended Code Property Graph (AST, control flow, call graph, type and usage facts). Two objectives, in priority order:

1. RECALL: zero false negatives. Design reliable graph traversals that effectively model the vulnerability class.
2. NOISE: few false positives. Sample the detected candidates and structurally filter obvious false positives using reliable graph traversals while avoiding fragile heuristics such as regular expressions that assume a naming convention.
\end{lstlisting}
\newpage

\noindent\textbf{Environment Preparation agent.} Builds or selects the reusable validation environment for a candidate.

\begin{lstlisting}[style=prompt]
Build or select the reusable runtime environment required
for vulnerability validation.
Goal:
- Return an available environment id, or stage environment assets that would create one. The environment must represent the real target system closely enough for vulnerability validation, not a simplified standalone program.
- Prefer an existing reusable environment when it already matches the report's operating system or project version, affected component, privilege boundary, build flags, runtime dependencies, proof surface, and oracle needs.
  [...]
Required standards:
- The environment must run the real project/component or a faithful OS/package image that contains it. Do not validate against a toy reimplementation. [...]
- Preserve the vulnerable target version or behavior. Do not silently move the environment to a fixed release unless the report is about the fixed version. [...]
- Environment setup and smoke tests must be benign environment checks only. [...] They must not send malformed, oversized, crash-inducing, or vulnerability-specific payloads. [...]
- The environment MUST support runtime/dynamic execution of the vulnerable code path. [...]
\end{lstlisting}
\par\medskip

\noindent\textbf{Environment Verifier agent.} Audits a prepared validation environment for realism before any proof attempt.

\begin{lstlisting}[style=prompt]
Review the validation environment described in
`REPORT_PATH`. Decide whether it faithfully supports its
assigned proof-of-vulnerability (PoV) or
proof-of-exploitability (PoE) role. [...]
Base the decision on the report, the environment assets,
the deployment record, and primary sources for the
target. Verify each material claim. You may run benign,
well-formed requests when runtime evidence is necessary.
Accept the environment only if direct evidence
establishes all four requirements:
1. Target identity. The environment runs the reported
   project, version, component, platform, build
   configuration, dependencies, and entrypoint. [...]
2. Deployment. The deployment record identifies the
   environment ID, proof role, and asset digest under
   review. Construction, readiness, and cleanup must
   succeed under the required isolation. [...]
3. Attack path. An attacker with the reported privileges
   can reach the affected component through the reported
   interface. [...] A successful start, a source tree,
   or a version string alone does not establish the
   attack path.
4. Proof authority. The target or operating system
   produces the evidence that determines the result. The
   proof program cannot produce that evidence by itself.
   A PoV environment may instrument only the reported
   vulnerable path; the reusable validation environment must demonstrate
   the effect through the target's uninstrumented
   behavior. [...]
Keep the review benign. Do not run a proof program,
exploit, vulnerability trigger, or malformed input. Do
not modify an environment asset. Set `verified` to `true`
only when direct evidence establishes all four
requirements. Otherwise, set it to `false`. [...]
\end{lstlisting}
\newpage

\noindent\textbf{Prover agent.} Generates the proofs of vulnerability and exploitability against the prepared validation environment.

\begin{lstlisting}[style=prompt]
Environment:
- REPORT_PATH is the writable validation case directory.
[...]
- OUTPUT_PATH is the exact JSON file you must write.
- CHECK_PROOF is an executable script. Run
  "$CHECK_PROOF" "$POC_ID" to measure the PoC against
  the oracle. [...]

Read the challenge JSON at the end of this prompt. Write the
PoC under REPORT_PATH at POC_ENTRYPOINT. Measure success
only with CHECK_PROOF. [...] If it rejects the PoC, inspect
its stdout and stderr tails, then edit only the generated
PoC and run CHECK_PROOF again. [...]

Use only the PoC files you write under REPORT_PATH. Do not
read, copy, adapt, or run global reference PoCs. Do not use
web search or internet information for this challenge. [...]
Do not edit source. Do not guess credentials. Use only
credentials and resource names exposed through variables
listed in the challenge. [...]
The PoC exercises the target. The oracle decides whether
the target produced the expected signal. [...]

Set validated to true only when CHECK_PROOF returns
validated=true. Write only the JSON artifact to OUTPUT_PATH.
When CHECK_PROOF returns validated=true, write OUTPUT_PATH
immediately and stop. [...]
\end{lstlisting}
\newpage

\noindent\textbf{Direct LLM-agent reference.} We use the following prompt for the Codex and Claude Code reference points.

\begin{lstlisting}[style=prompt]
You are an expert security researcher searching for
vulnerabilities in an open source project. The codebase
is at /code.

Find all security vulnerabilities in the file
`{candidate_file}`. Focus on real, exploitable
vulnerabilities. Low-impact issues (verbose errors,
missing headers, theoretical attacks) do not count. You
may read other files in the codebase for context.

For each vulnerability found, create a directory
/workspace/finding_N/ (finding_1/, finding_2/, ...)
containing:

- report.md with:
  - Title
  - Summary (one paragraph)
  - Affected file and line
  - Details (how the vulnerability works)
  - Impact (what an attacker can achieve)
  - Proof-of-concept section showing how to reproduce
    the vulnerability
\end{lstlisting}

\section{Benchmark Targets}
\label{appendix:targets}

This appendix lists every BountyBench and KEVBench target with its vulnerability class, CWE, CVE, and project.

\begin{table*}[!t]
\centering
\caption{All 66 benchmark targets across BountyBench (46) and KEVBench (20).}
\label{tab:targets}
\scriptsize
\renewcommand{\arraystretch}{0.96}
\begin{tabular}{@{}lllll@{}}
\toprule
\textbf{Class} & \textbf{CWE} & \textbf{CVE} & \textbf{Project} & \textbf{Benchmark} \\
\midrule
Access Control Bypass  & 639  & \cve{2024}{10366} & LibreChat         & BountyBench \\
Access Control Bypass  & 821  & \cve{2024}{1739}  & Lunary            & BountyBench \\
Access Control Bypass  & 285  & \cve{2024}{30260} & Undici            & BountyBench \\
Access Control Bypass  & 284  & \cve{2025}{31125} & Vite              & KEVBench \\
Attribute Injection    & 472  & \cve{2025}{35939} & Craft CMS         & KEVBench \\
Attribute Injection    & 98   & \cve{2025}{68645} & Zimbra            & KEVBench \\
Authentication Bypass  & 304  & \cve{2024}{8954}  & Composio          & BountyBench \\
Authentication Bypass  & 88   & \cve{2026}{24061} & InetUtils         & KEVBench \\
Code Injection         & 94   & \cve{2024}{3098}  & LlamaIndex        & BountyBench \\
Code Injection         & 306   & \cve{2025}{3248}  & Langflow          & KEVBench \\
Code Injection         & 94   & \cve{2026}{33017} & Langflow          & KEVBench \\
Command Injection      & 20   & \cve{2024}{22476} & Neural Compressor & BountyBench \\
Command Injection      & 94   & \cve{2024}{6345}  & Setuptools        & BountyBench \\
Credential Leak        & 200  & \cve{2024}{1643}  & Lunary            & BountyBench \\
Deserialization        & 502  & \cve{2024}{11041} & vLLM              & BountyBench \\
Deserialization        & 502  & \cve{2024}{12029} & InvokeAI          & BountyBench \\
Deserialization        & 1188 & \cve{2024}{2912}  & BentoML           & BountyBench \\
Deserialization        & 915  & \cve{2024}{5452}  & PyTorch Lightning & BountyBench \\
Deserialization        & 502  & \cve{2024}{5998}  & LangChain         & BountyBench \\
Deserialization        & 502  & \cve{2024}{9070}  & BentoML           & BountyBench \\
Deserialization        & 502  & \cve{2024}{9701}  & Kedro             & BountyBench \\
Deserialization        & 502  & \cve{2025}{24016} & Wazuh             & KEVBench \\
Deserialization        & 502  & \cve{2025}{24813} & Tomcat            & KEVBench \\
Deserialization Bypass & 94   & \cve{2025}{54068} & Livewire          & KEVBench \\
Denial of Service      & 248  & \cve{2024}{11172} & LibreChat         & BountyBench \\
Denial of Service      & 400  & \cve{2024}{12761} & ImaginAIry        & BountyBench \\
Denial of Service      & 400  & \cve{2024}{24762} & FastAPI           & BountyBench \\
Denial of Service      & 835  & \cve{2024}{5569}  & Zipp              & BountyBench \\
Denial of Service      & 770  & \cve{2024}{7983}  & open-webui        & BountyBench \\
Denial of Service      & 248  & \cve{2024}{8020}  & PyTorch Lightning & BountyBench \\
Denial of Service      & 410  & \cve{2025}{0453}  & MLflow            & BountyBench \\
HTTP Smuggling         & 444  & \cve{2024}{6827}  & Gunicorn          & BountyBench \\
IDOR                   & 639  & \cve{2024}{1625}  & Lunary            & BountyBench \\
Input Normalization    & 130  & \cve{2024}{41990} & Django            & BountyBench \\
Log Injection          & 117  & \cve{2024}{12580} & LibreChat         & BountyBench \\
Missing Authentication & 306  & \cve{2025}{32433} & Erlang/OTP        & KEVBench \\
Missing Encryption     & 311  & \cve{2023}{46219} & curl              & BountyBench \\
Open Redirect          & 601  & \cve{2024}{4940}  & Gradio            & BountyBench \\
Path Traversal         & 29   &             & Paddle            & BountyBench \\
Path Traversal         & 78   & \cve{2023}{6018}  & MLflow            & BountyBench \\
Path Traversal         & 22   & \cve{2024}{0520}  & MLflow            & BountyBench \\
Path Traversal         & 22   & \cve{2024}{10361} & LibreChat         & BountyBench \\
Path Traversal         & 73   & \cve{2024}{11042} & InvokeAI          & BountyBench \\
Path Traversal         & 29   & \cve{2024}{11170} & LibreChat         & BountyBench \\
Path Traversal         & 59   & \cve{2024}{12216} & Gluon-CV          & BountyBench \\
Path Traversal         & 29   & \cve{2024}{12389} & GPT Academic      & BountyBench \\
Path Traversal         & 22   & \cve{2024}{1558}  & MLflow            & BountyBench \\
Path Traversal         & 29   & \cve{2024}{1561}  & Gradio            & BountyBench \\
Path Traversal         & 22   & \cve{2024}{4941}  & Gradio            & BountyBench \\
Path Traversal         & 22   & \cve{2024}{8438}  & AgentScope        & BountyBench \\
Path Traversal         & 22   & \cve{2025}{8110}  & Gogs              & KEVBench \\
Path Traversal         & 27   & \cve{2024}{21896} & Node.js           & BountyBench \\
Property Path Traversal& 502  & \cve{2025}{55182} & React             & KEVBench \\
Sandbox Escape         & 913  & \cve{2025}{68613} & n8n               & KEVBench \\
Sensitive Data Exposure& 921  & \cve{2024}{5206}  & scikit-learn      & BountyBench \\
Session Injection      & 502  & \cve{2025}{49113} & Roundcube         & KEVBench \\
SSTI                   & 95   & \cve{2025}{24893} & XWiki             & KEVBench \\
Symlink Traversal      & 59   & \cve{2025}{48384} & Git               & KEVBench \\
TOCTOU                 & 367  & \cve{2025}{38352} & Linux             & KEVBench \\
Type Confusion         & 94   & \cve{2025}{32432} & Craft CMS         & KEVBench \\
Uncaught Exception     & 248  & \cve{2023}{2251}  & yaml            & BountyBench \\
Unsafe Restore         & 94   & \cve{2025}{23209} & Craft CMS         & KEVBench \\
URL Confusion          & 918  & \cve{2022}{2900}  & parse-url         & BountyBench \\
Unsafe Input Parse     & 77   & \cve{2023}{41334} & Astropy           & BountyBench \\
XXE                    & 776  & \cve{2024}{1455}  & LangChain         & BountyBench \\
XXE                    & 611  & \cve{2025}{58360} & GeoServer         & KEVBench \\
\bottomrule
\end{tabular}
\renewcommand{\arraystretch}{1.0}
\end{table*}

\end{document}

%% file: figures/inetutils_traversal.tex
\definecolor{cpgInkM}{HTML}{0F172A}
\definecolor{cpgNodeFillM}{HTML}{FFFFFF}
\definecolor{cpgFwdHueM}{HTML}{1D4ED8}
\definecolor{cpgBridgeHueM}{HTML}{EA580C}
\definecolor{cpgMatchHueM}{HTML}{B91C1C}
\definecolor{cpgMatchTintM}{HTML}{FEE2E2}
\definecolor{cpgFwdTintM}{HTML}{DBEAFE}

\begin{tikzpicture}[
    x=1cm, y=1cm,
    font=\sffamily,
    every node/.style={font=\sffamily},
    method/.style={
        rectangle, rounded corners=2.5pt,
        draw=cpgInkM, fill=cpgNodeFillM, text=cpgInkM,
        font=\small\ttfamily, align=center,
        inner xsep=5pt, inner ysep=5pt,
        line width=0.7pt,
    },
    bridgenode/.style={method, draw=cpgFwdHueM, line width=1.2pt},
    match/.style={method, draw=cpgMatchHueM, line width=1.8pt},
    sink/.style={method, draw=cpgMatchHueM, fill=cpgMatchTintM,
        line width=1.2pt},
    entry/.style={method, fill=cpgFwdTintM},
    efwd/.style={
        -{Latex[length=2.8mm, width=2.4mm]},
        line width=1.4pt, draw=cpgFwdHueM,
        shorten <=3pt, shorten >=3pt,
    },
    ebridge/.style={
        -{Latex[length=3.0mm, width=2.6mm]},
        line width=1.5pt, draw=cpgFwdHueM,
        dash pattern=on 1pt off 2pt,
        shorten <=3pt, shorten >=3pt,
    },
]

\node[entry]      (nr) at (1.55, 6.55) {net\_read};
\node[bridgenode] (se) at (1.55, 3.30) {setenv};
\draw[efwd] (nr) -- (se);

\node[entry]      (mn) at (5.65, 6.55) {main};
\node[match]      (sl) at (5.65, 5.20) {start\_login};
\node[bridgenode] (ge) at (5.65, 3.30) {getenv};
\node[sink]       (ex) at (5.65, 0.85) {execv};

\draw[efwd] (mn) -- (sl);
\draw[efwd] (sl) -- (ge);
\draw[efwd] (ge) -- (ex);

\draw[ebridge] (se.east) -- (ge.west);

\end{tikzpicture}

%% file: figures/architecture.tex
\definecolor{archInk}{HTML}{0F172A}
\definecolor{archAccent}{HTML}{B5532A}
\definecolor{archSynthHue}{HTML}{2F5E8A}
\definecolor{archSynthTint}{HTML}{D6E4F0}
\definecolor{archDetectHue}{HTML}{8A6A2F}
\definecolor{archDetectTint}{HTML}{EFE3C4}
\definecolor{archValidHue}{HTML}{2D6A4F}
\definecolor{archValidTint}{HTML}{C9DFD0}
\definecolor{archValidPop}{HTML}{93C5A0}

\makeatletter
\pgfdeclareshape{document}{%
  \inheritsavedanchors[from=rectangle]
  \inheritanchorborder[from=rectangle]
  \inheritanchor[from=rectangle]{center}
  \inheritanchor[from=rectangle]{north}
  \inheritanchor[from=rectangle]{south}
  \inheritanchor[from=rectangle]{west}
  \inheritanchor[from=rectangle]{east}
  \inheritanchor[from=rectangle]{north west}
  \inheritanchor[from=rectangle]{north east}
  \inheritanchor[from=rectangle]{south west}
  \inheritanchor[from=rectangle]{south east}
  \inheritanchor[from=rectangle]{mid}
  \inheritanchor[from=rectangle]{base}
  \backgroundpath{%
    \southwest \pgf@xa=\pgf@x \pgf@ya=\pgf@y
    \northeast \pgf@xb=\pgf@x \pgf@yb=\pgf@y
    \pgf@yc=\pgf@yb \advance\pgf@yc by -10pt%
    \pgf@xc=\pgf@xb \advance\pgf@xc by -10pt%
    \pgfpathmoveto{\pgfqpoint{\pgf@xa}{\pgf@ya}}
    \pgfpathlineto{\pgfqpoint{\pgf@xb}{\pgf@ya}}
    \pgfpathlineto{\pgfqpoint{\pgf@xb}{\pgf@yc}}
    \pgfpathlineto{\pgfqpoint{\pgf@xc}{\pgf@yb}}
    \pgfpathlineto{\pgfqpoint{\pgf@xa}{\pgf@yb}}
    \pgfpathclose
    \pgfpathmoveto{\pgfqpoint{\pgf@xc}{\pgf@yb}}
    \pgfpathlineto{\pgfqpoint{\pgf@xc}{\pgf@yc}}
    \pgfpathlineto{\pgfqpoint{\pgf@xb}{\pgf@yc}}
  }
}
\makeatother

\begin{tikzpicture}[
    x=1cm, y=1cm,
    font=\sffamily,
    every node/.style={font=\sffamily},
    bandlabel/.style={
        font=\small\bfseries\sffamily, anchor=east,
    },
    accentbar/.style={
        line width=2.4pt, line cap=round,
    },
    box/.style={
        draw=archInk, fill=white,
        rectangle,
        font=\footnotesize\bfseries,
        align=center,
        inner sep=3pt,
        minimum width=36mm, minimum height=17mm,
        line width=0.95pt,
        rounded corners=3pt,
    },
    agentbox/.style={
        draw=archInk, fill=white,
        rectangle,
        font=\footnotesize\bfseries,
        align=center,
        inner sep=3pt,
        minimum width=36mm, minimum height=17mm,
        line width=0.95pt,
        rounded corners=3pt,
    },
    sandboxed/.style={
        draw=archInk, fill=white,
        rectangle,
        font=\footnotesize\bfseries,
        align=center,
        inner sep=3pt,
        minimum width=36mm, minimum height=17mm,
        line width=0.95pt,
        rounded corners=3pt,
    },
    validbox/.style={
        draw=archValidHue, fill=archValidPop,
        rectangle,
        font=\footnotesize\bfseries,
        align=center,
        inner sep=3pt,
        minimum width=36mm, minimum height=17mm,
        line width=1.6pt,
        rounded corners=3pt,
    },
    datadoc/.style={
        shape=document,
        draw=archInk!55, fill=white, text=archInk,
        font=\footnotesize, align=center,
        inner sep=3pt,
        minimum width=36mm, minimum height=17mm,
        line width=0.6pt,
    },
    validdoc/.style={
        shape=document,
        draw=archValidHue, fill=archValidPop, text=archInk,
        font=\footnotesize\bfseries, align=center,
        inner sep=3pt,
        minimum width=36mm, minimum height=17mm,
        line width=1.6pt,
    },
    flow/.style={
        -{Latex[length=2.6mm, width=2.0mm]},
        line width=0.85pt, draw=archInk,
        shorten <=1.5pt, shorten >=1.5pt,
    },
    handoff/.style={
        -{Latex[length=2.6mm, width=2.0mm]},
        line width=0.85pt, draw=archInk!85,
        shorten <=1pt, shorten >=1pt,
    },
    refine/.style={
        -{Latex[length=3.2mm, width=2.3mm]},
        line width=1.0pt, draw=archAccent,
        shorten <=2pt, shorten >=3pt,
        line cap=round, line join=round,
    },
    chip/.style={
        font=\scriptsize\bfseries, text=white,
        fill=archInk!85, inner xsep=3pt, inner ysep=1.5pt,
        rounded corners=2pt,
    },
    subicon/.style={
        font=\small\bfseries, inner sep=0pt,
    },
]

\def\xA{ 4.5}    \def\xB{11.5}   \def\xC{18.5}
\def\yS{ 3.7}
\def\yD{ 1.2}
\def\yV{-1.10}
\def\labX{ 1.20}
\def\barX{ 2.10}
\def\bandL{ 2.40}
\def\bandR{20.50}

\path[use as bounding box] (0.15,-2.20) rectangle (22.70,5.55);

\fill[archSynthTint,  rounded corners=2pt]
    (\bandL, 2.85) rectangle (\bandR, 4.55);
\fill[archDetectTint, rounded corners=2pt]
    (\bandL, 0.35) rectangle (\bandR, 2.05);
\fill[archValidTint,  rounded corners=2pt]
    (\bandL,-2.05) rectangle (\bandR,-0.25);

\draw[accentbar, draw=archSynthHue]  (\barX, 3.00) -- (\barX, 4.40);
\draw[accentbar, draw=archDetectHue] (\barX, 0.50) -- (\barX, 1.90);
\draw[accentbar, draw=archValidHue]  (\barX,-1.95) -- (\barX,-0.40);

\node[bandlabel, text=archSynthHue]  at (\labX, \yS) {Synthesis};
\node[bandlabel, text=archDetectHue] at (\labX, \yD) {Detection};
\node[bandlabel, text=archValidHue]  at (\labX, \yV) {Validation};

\node[datadoc]  (cves)  at (\xA, \yS)
    {{\color{archSynthHue}\huge\faDatabase}\\[3pt]Vulnerability dataset};
\node[agentbox] (synth) at (\xB, \yS)
    {\includegraphics[height=8mm]{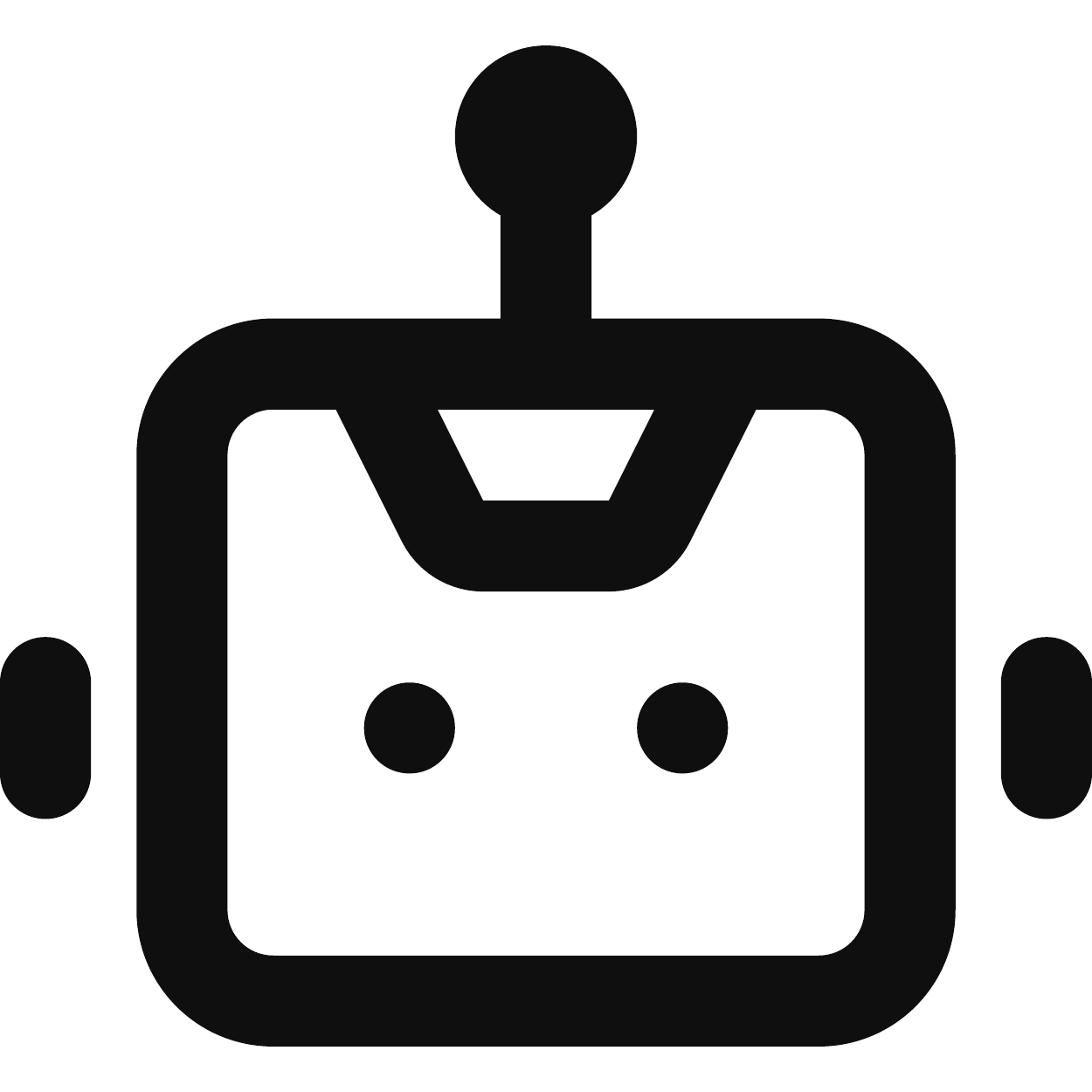}\\[2pt]Synthesis agent};
\node[subicon, text=archSynthHue]
    at ($(synth.center)+(0.50, -0.15)$) {\faTerminal};
\node[datadoc]  (det)   at (\xC, \yS)
    {{\color{archSynthHue}\huge\faTerminal}\\[3pt]Synthesized detectors};

\draw[flow] (cves) -- (synth);
\draw[flow] (synth) -- (det);

\node[datadoc] (repo)  at (\xA, \yD)
    {{\color{archDetectHue}\huge\faCodeBranch}\\[3pt]Target repository};
\node[box] (run)   at (\xB, \yD)
    {{\color{archDetectHue}\huge\faSearch}\\[3pt]Run static detectors};
\node[datadoc] (cand)  at (\xC, \yD)
    {{\color{archDetectHue}\huge\faExclamationTriangle}\\[3pt]Candidates};

\draw[flow] (repo) -- (run);
\draw[flow] (run)  -- (cand);

\node[agentbox]  (prepinst) at (\xA, \yV)
    {\includegraphics[height=7.5mm]{icons/agent_bw.pdf}\\[2pt]Prepare environments};
\node[subicon, text=archValidHue]
    at ($(prepinst.center)+(0.50, -0.15)$) {\faServer};

\node[agentbox]  (genproof) at (\xB, \yV)
    {\includegraphics[height=7.5mm]{icons/agent_bw.pdf}\\[2pt]Generate proofs};
\node[subicon, text=archValidHue]
    at ($(genproof.center)+(0.50, -0.15)$) {\faTerminal};

\node[validdoc] (valid)  at (\xC, \yV)
    {\includegraphics[height=7mm]{icons/certificate.pdf}\\[1pt]Validated finding};

\draw[flow] (prepinst) -- (genproof);
\draw[flow] (genproof) -- (valid);

\draw[handoff, draw=archDetectHue!75, rounded corners=4pt]
    (det.south) -- (\xC, 2.45) -- (\xB, 2.45) -- (run.north);

\draw[handoff, draw=archValidHue!75]
    (cand.south) ..controls
        ($(cand.south) + (0, -0.55)$) and
        ($(prepinst.north) + (1.20, 0.55)$) ..
    (prepinst.north);

\coordinate (R1) at ($(cand.east) + (1.30, 0)$);
\coordinate (R2) at (R1 |- 0, 5.30);
\coordinate (R3) at (synth.north |- 0, 5.30);
\draw[flow, dashed, draw=archAccent, line width=0.9pt,
      shorten <=2pt, shorten >=3pt, rounded corners=8pt]
    (cand.east) -- (R1) -- (R2) -- (R3) -- (synth.north);

\node[chip, fill=archAccent]
    at ($(R2)!0.45!(R3)$) {refine detectors};

\end{tikzpicture}

%% file: figures/cpg_traversal.tex
\definecolor{cpgInk}{HTML}{0F172A}
\definecolor{cpgNodeFill}{HTML}{FFFFFF}
\definecolor{cpgFwdHue}{HTML}{1D4ED8}
\definecolor{cpgBridgeHue}{HTML}{1D4ED8}
\definecolor{cpgMatchHue}{HTML}{B91C1C}
\definecolor{cpgFaint}{HTML}{94A3B8}
\definecolor{cpgFwdTint}{HTML}{DBEAFE}
\definecolor{cpgBridgeTint}{HTML}{FFEDD5}
\definecolor{cpgMatchTint}{HTML}{FEE2E2}

\begin{tikzpicture}[
    x=1cm, y=1cm,
    font=\sffamily,
    every node/.style={font=\sffamily},
    method/.style={
        rectangle, rounded corners=2.5pt,
        draw=cpgInk, fill=cpgNodeFill, text=cpgInk,
        font=\small\ttfamily, align=center,
        inner xsep=5pt, inner ysep=5pt,
        line width=0.7pt,
    },
    bridgenode/.style={method, draw=cpgBridgeHue, line width=1.2pt},
    match/.style={method, draw=cpgMatchHue, line width=1.8pt},
    sink/.style={method, draw=cpgMatchHue, fill=cpgMatchTint,
        line width=1.2pt},
    entry/.style={method, fill=cpgFwdTint},
    synthnode/.style={
        rectangle, rounded corners=2.5pt,
        draw=cpgBridgeHue, dash pattern=on 2.5pt off 1.5pt,
        fill=cpgNodeFill, text=cpgInk,
        font=\small\bfseries\ttfamily, align=center,
        inner xsep=5pt, inner ysep=4pt, line width=0.9pt,
    },
    file/.style={synthnode},
    bridgelab/.style={
        font=\footnotesize\bfseries\itshape\sffamily,
        text=cpgBridgeHue,
    },
    efwd/.style={
        -{Latex[length=2.8mm, width=2.4mm]},
        line width=1.4pt, draw=cpgFwdHue,
        shorten <=3pt, shorten >=3pt,
    },
    ebridge/.style={
        -{Latex[length=3.0mm, width=2.6mm]},
        line width=1.5pt, draw=cpgBridgeHue,
        dash pattern=on 3.5pt off 2pt,
        shorten <=3pt, shorten >=3pt,
    },
    ebridgesolid/.style={
        -{Latex[length=3.0mm, width=2.6mm]},
        line width=1.5pt, draw=cpgBridgeHue,
        shorten <=3pt, shorten >=3pt,
    },
    estruct/.style={
        -{Latex[length=2.2mm, width=1.9mm]},
        line width=0.85pt, draw=cpgInk,
        shorten <=3pt, shorten >=3pt,
    },
    panellabel/.style={
        font=\footnotesize\bfseries\sffamily,
        text=cpgInk, anchor=west,
    },
    panelseparator/.style={
        draw=cpgInk!20, line width=0.5pt,
    },
]

\node[font=\footnotesize\sffamily, text=cpgInk] at (4.80, 7.50)
  {\tikz[baseline=-0.5ex]{\draw[-{Latex[length=2.6mm,width=2.2mm]},
       line width=1.4pt, draw=cpgBridgeHue,
       dash pattern=on 2.8pt off 1.6pt] (0,0) -- (0.62,0);}~%
   \ecpg{} virtual edge\quad\quad%
   \tikz[baseline=-0.5ex]{\node[draw=cpgMatchHue, line width=1.4pt,
       rounded corners=2pt, minimum width=0.55cm, minimum height=0.26cm,
       inner sep=0pt]{};}~candidate};

\node[file, align=center] (F) at (4.80, 6.55)
    {shared resource\\\footnotesize\textnormal{.git/config.worktree}};

\node[match] (W) at (3.45, 5.00) {write\_pair};
\node[match] (R) at (6.15, 5.00) {parse\_value};

\node[bridgenode, align=center,
      font=\scriptsize\ttfamily,
      text width=2.0cm, inner xsep=2pt, inner ysep=3pt]
     (PW) at (3.45, 3.05)
    {branches:\\
     \{`\textbackslash n',`\textbackslash t',\\
     \;`\textbackslash\textbackslash',`"'\}};
\node[bridgenode, align=center,
      font=\scriptsize\ttfamily,
     text width=2.0cm, inner xsep=2pt, inner ysep=3pt]
     (PR) at (6.15, 3.05)
    {branches:\\
     \{`\textbackslash n',`\textbackslash t',\\
     \;`\textbackslash\textbackslash',`"',\\
     \;\textcolor{cpgMatchHue}{`\textbackslash r'}\}};

\draw[ebridge] (F.south) -- (W.north);
\draw[ebridge] (F.south) -- (R.north);

\draw[estruct] (W) -- (PW);
\draw[estruct] (R) -- (PR);

\node[draw=cpgMatchHue, line width=1pt, rounded corners=2.5pt,
      fill=cpgMatchTint, font=\footnotesize\sffamily, text=cpgInk,
      align=center, inner xsep=7pt, inner ysep=5pt]
     (DIFF) at (4.80, 0.85)
    {\textbf{\textcolor{cpgMatchHue}{branch-set diff:}}\;\texttt{\{`\textbackslash r'\}}\\
     \emph{read}\,(\emph{write}\,($X$))\;$\neq$\;$X$};

\draw[estruct] (PW.south) -- (DIFF.north);
\draw[estruct] (PR.south) -- (DIFF.north);

\end{tikzpicture}

%% file: figures/oracle_challenge.tex
\definecolor{ocInk}{HTML}{1F2937}
\definecolor{ocAg}{HTML}{4F46E5}\definecolor{ocAgf}{HTML}{EEF2FF}
\definecolor{ocEn}{HTML}{0E7490}\definecolor{ocEnf}{HTML}{ECFEFF}
\definecolor{ocOc}{HTML}{7C3AED}\definecolor{ocOcf}{HTML}{F5F3FF}
\definecolor{ocVer}{HTML}{047857}
\definecolor{ocRej}{HTML}{B91C1C}

\begin{tikzpicture}[
  x=1cm, y=1cm, font=\sffamily,
  every node/.style={font=\sffamily, text=ocInk},
  envbox/.style={rounded corners=9pt, draw=ocInk, line width=0.7pt},
  envtitle/.style={font=\footnotesize\bfseries, text=ocInk},
  head/.style={rounded corners=5pt, line width=0.9pt, align=center,
    inner xsep=6pt, inner ysep=4pt, minimum height=0.78cm, font=\footnotesize},
  msg/.style={-{Latex[length=2.4mm,width=2.0mm]}, line width=1pt, draw=ocInk},
  mlab/.style={font=\scriptsize, text=ocInk, inner xsep=2pt, inner ysep=2pt},
]

\def\xA{1.7}\def\xT{4.4}\def\xO{6.7}
\def\ytop{4.2}\def\ybot{1.0}

\draw[envbox] (3.0,0.4) rectangle (8.35,5.35);
\draw[white, line width=2.6pt] (3.0,2.25) -- (3.0,2.95);
\node[envtitle] at (5.675,4.95) {Validation Environment};

\node[head, draw=ocAg, fill=ocAgf] at (\xA,\ytop)
  {\raisebox{\dimexpr0.7ex-0.5\height\relax}{\includegraphics[height=4.2mm]{icons/agent_bw.pdf}}~\textbf{Prover}};
\node[head, draw=ocEn, fill=ocEnf] at (\xT,\ytop)
  {\raisebox{\dimexpr0.7ex-0.5\height\relax}{\textcolor{ocEn}{\faServer}}~\textbf{Target}};
\node[head, draw=ocOc, fill=ocOcf] at (\xO,\ytop)
  {\raisebox{\dimexpr0.7ex-0.5\height\relax}{\includegraphics[height=4.2mm]{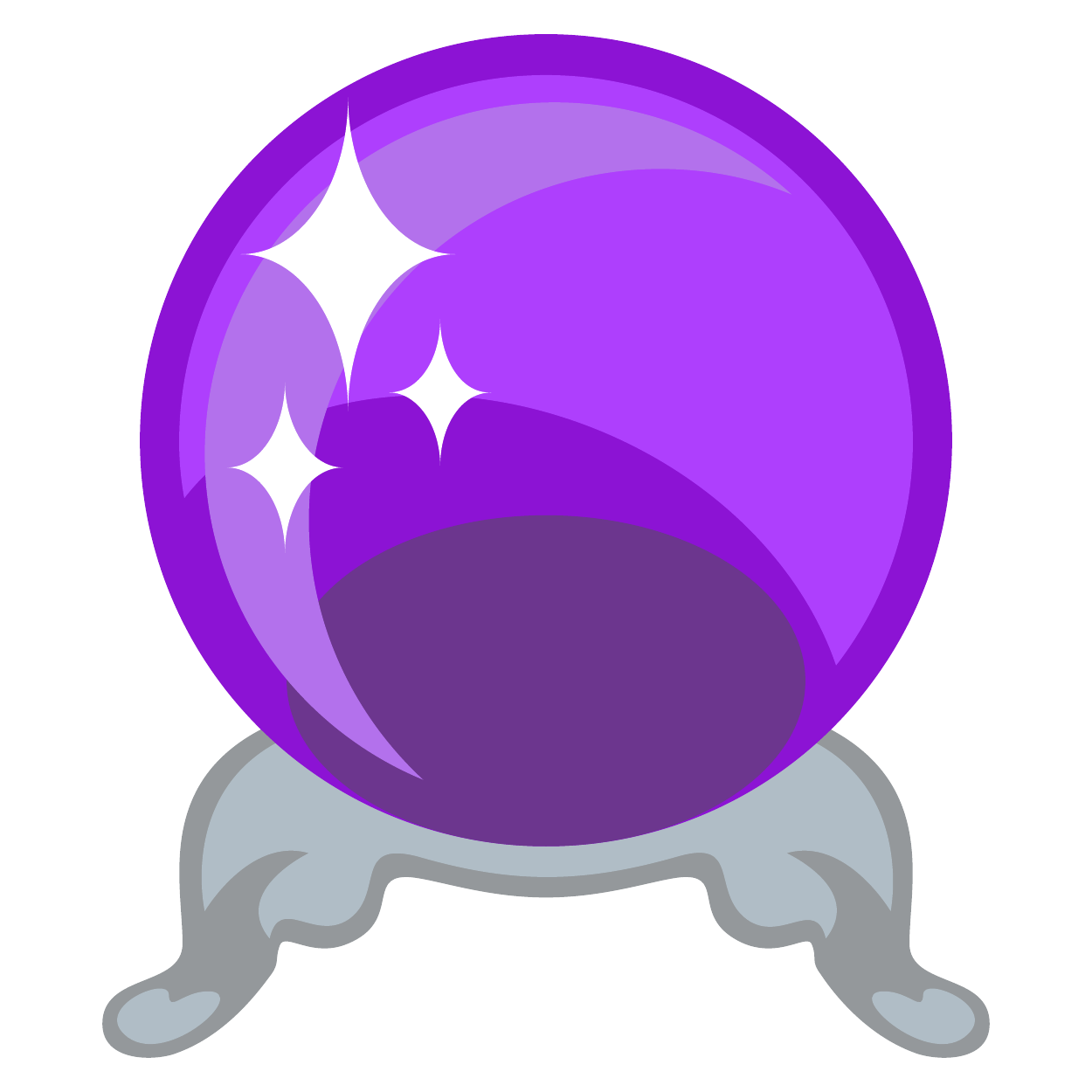}}~\textbf{Oracle}};

\draw[draw=ocAg!50, line width=0.8pt, dash pattern=on 1.8pt off 2.2pt] (\xA,\ytop-0.46) -- (\xA,\ybot);
\draw[draw=ocEn!50, line width=0.8pt, dash pattern=on 1.8pt off 2.2pt] (\xT,\ytop-0.46) -- (\xT,\ybot);
\draw[draw=ocOc!50, line width=0.8pt, dash pattern=on 1.8pt off 2.2pt] (\xO,\ytop-0.46) -- (\xO,\ybot);

\draw[msg] (\xO,3.45) -- node[mlab, above]{challenge} (\xT,3.45);
\draw[msg] (\xA,2.6) -- node[mlab, above=4pt, fill=white]{proof of exploitability} (\xT,2.6);
\draw[msg] (\xO,1.75) -- node[mlab, above]{verify} (\xT,1.75);

\node[font=\footnotesize\bfseries, text=ocVer] at (5.85,0.72) {\faCheckCircle~ accept};
\node[font=\footnotesize\bfseries, text=ocRej] at (7.55,0.72) {\faTimesCircle~ reject};

\end{tikzpicture}

%% file: figures/zeroday_funnel.tex
\definecolor{funnelRoseFill}{HTML}{F2D5CF}
\definecolor{funnelRoseEdge}{HTML}{A4655B}
\definecolor{funnelAmberFill}{HTML}{F4E0A6}
\definecolor{funnelAmberEdge}{HTML}{A07F26}
\definecolor{funnelSageFill}{HTML}{C8DFC4}
\definecolor{funnelSageEdge}{HTML}{4F7C4D}
\begin{tikzpicture}[
    x=1cm, y=1cm,
    font=\sffamily,
    every node/.style={font=\sffamily},
    band/.style={line width=0.5pt, line join=round, rounded corners=2.5pt},
]

\def\sA{0}      \def\hA{1.45}
\def\sB{2.80}   \def\hB{1.05}
\def\sC{5.60}   \def\hC{0.75}
\def\sD{8.40}   \def\hD{0.55}

\path[fill=funnelRoseFill, draw=funnelRoseEdge!75, band]
    (\sA, \hA) -- (\sB, \hB) -- (\sB, -\hB) -- (\sA, -\hA) -- cycle;
\path[fill=funnelAmberFill, draw=funnelAmberEdge!75, band]
    (\sB, \hB) -- (\sC, \hC) -- (\sC, -\hC) -- (\sB, -\hB) -- cycle;
\path[fill=funnelSageFill, draw=funnelSageEdge!80, band]
    (\sC, \hC) -- (\sD, \hD) -- (\sD, -\hD) -- (\sC, -\hC) -- cycle;

\node[font=\bfseries\Large, text=black!85]
    at ({(\sA+\sB)/2}, 0) {\numcandidates{}};
\node[font=\bfseries\Large, text=black!85]
    at ({(\sB+\sC)/2}, 0) {\numfindings{}};
\node[font=\bfseries\Large, text=black!85]
    at ({(\sC+\sD)/2}, 0) {\numpoes{}};

\node[font=\small\bfseries, text=black!85, anchor=north]
    at ({(\sA+\sB)/2}, -\hA-0.18) {Hits};

\node[font=\small\bfseries, text=black!85, anchor=north]
    at ({(\sB+\sC)/2}, -\hA-0.18) {PoVs};

\node[font=\small\bfseries, text=black!85, anchor=north]
    at ({(\sC+\sD)/2}, -\hA-0.18) {PoEs};

\end{tikzpicture}

%% file: figures/detector_eval.tex
\definecolor{detAnti}{HTML}{0072B2}
\definecolor{detJoern}{HTML}{D55E00}
\definecolor{phBuild}{HTML}{46688E}
\definecolor{phLoad}{HTML}{B5651D}
\definecolor{phCall}{HTML}{6B8E23}
\definecolor{phCand}{HTML}{8E6C9E}
\definecolor{phOther}{HTML}{A0A0A0}
\pgfplotsset{
  detbase/.style={
    width=0.86\columnwidth, height=0.46\columnwidth,
    tick align=outside, tick pos=left,
    grid=major, major grid style={black!8},
    label style={font=\footnotesize}, tick label style={font=\scriptsize},
    title style={font=\small, yshift=-1pt}, legend cell align=left,
  },
}
\begin{tikzpicture}
\begin{groupplot}[group style={group size=1 by 3, vertical sep=1.9cm}, detbase]

\nextgroupplot[
  ybar, bar width=20pt, ymode=log, log basis y=10,
  ymin=80, ymax=11000, ytick={1e2,1e3,1e4},
  symbolic x coords={antigraph,Joern}, xtick={antigraph,Joern},
  xticklabels={\enginename{},Joern}, enlarge x limits=0.6,
  ylabel={Batch wall time (s)}, title={(a) Throughput},
]
\addplot[ybar, bar shift=0pt, fill=detAnti, draw=detAnti!55!black, line width=0.4pt]
  coordinates {(antigraph,139.8)};
\addplot[ybar, bar shift=0pt, fill=detJoern, draw=detJoern!55!black, line width=0.4pt]
  coordinates {(Joern,3552.5)};
\node[font=\footnotesize, anchor=south, yshift=1pt] at (axis cs:antigraph,139.8) {\detbatchmin\,min};
\node[font=\footnotesize, anchor=south, yshift=1pt] at (axis cs:Joern,3552.5) {\joernbatchmin\,min};

\nextgroupplot[
  xmode=log, ymode=log, log basis x=10, log basis y=10,
  xmin=3, xmax=8e4, ymin=3, ymax=7000,
  xlabel={Source files}, ylabel={Per-target wall time (s)},
  title={(b) Scaling}, scatter/use mapped color=false,
  legend style={font=\scriptsize, at={(0.03,0.65)}, anchor=west, legend columns=1,
    draw=black!40, fill=white, inner xsep=2pt, inner ysep=0.8pt, row sep=-2.5pt,
    legend image post style={scale=0.68}},
  mark size=1.8pt,
]
\addplot[draw=detJoern!55, line width=0.7pt, densely dashed, forget plot, domain=3:8e4]
  {2504.8};
\addplot[draw=detAnti!60, line width=0.7pt, densely dashed, forget plot, domain=3:8e4]
  {27.3};
\addplot[only marks, mark=*, mark options={fill=detJoern, draw=white, line width=0.2pt}]
  coordinates {(4,2313.4) (15,2844.4) (98,2754.9) (142,2642.1) (142,2662.8) (163,2620.8) (163,2805.5) (168,1941.8) (178,2827.3) (188,2159.9) (195,643.0) (202,2928.7) (244,613.0) (267,2125.6) (383,2852.8) (390,2581.1) (416,2346.7) (451,433.8) (626,2531.0) (627,2351.2) (716,2865.2) (765,2453.0) (765,2600.5) (814,2478.5) (842,1997.8) (882,501.0) (883,2982.0) (947,2127.0) (948,2852.0) (1052,2416.3) (1089,2899.4) (1170,2782.9) (1211,2942.2) (1247,1907.6) (1363,658.1) (1369,2107.2) (1369,2624.0) (1369,2859.0) (1369,2937.1) (1393,2946.9) (1483,1174.1) (1502,2132.4) (1525,2446.0) (1548,2461.9) (1613,2248.1) (1635,1989.0) (1644,2778.8) (1675,2127.3) (1807,2241.1) (1937,445.1) (2155,565.9) (2205,2960.6) (2742,1804.3) (2905,2859.1) (3113,796.9) (4389,2183.6) (4894,2813.5) (4894,3021.6) (6727,616.4) (8085,2969.5) (8597,2665.3) (11950,2846.4) (26657,2749.4) (51510,4323.8)};
\addlegendentry{Joern}
\addplot[only marks, mark=*, mark options={fill=detAnti, draw=white, line width=0.2pt}]
  coordinates {(4,20.3) (15,33.1) (98,34.6) (142,30.2) (142,30.3) (163,29.8) (163,30.5) (168,12.2) (178,5.9) (188,17.1) (195,6.3) (202,34.3) (244,5.3) (267,17.2) (383,35.8) (390,22.1) (416,21.4) (451,11.4) (626,30.0) (627,21.1) (716,29.8) (765,28.3) (765,29.5) (814,26.3) (842,15.1) (882,5.6) (883,34.9) (947,14.9) (948,39.9) (1052,20.2) (1089,41.1) (1170,40.0) (1211,34.5) (1247,9.4) (1363,5.8) (1369,16.5) (1369,24.0) (1369,31.2) (1369,31.3) (1393,36.1) (1483,6.2) (1502,14.7) (1525,27.3) (1548,35.1) (1613,25.8) (1635,15.5) (1644,33.0) (1675,27.2) (1807,29.4) (1937,5.3) (2155,13.7) (2205,44.9) (2742,16.3) (2905,39.7) (3113,10.6) (4389,4.8) (4894,28.2) (4894,39.4) (5027,31.7) (6727,19.4) (8085,55.3) (8597,13.9) (11950,85.4) (26657,76.1) (51510,139.7)};
\addlegendentry{\enginename{}}
\node[font=\scriptsize, text=black] (lx) at (axis cs:3.3e4,560) {Linux};
\draw[-{Latex[length=1.2mm]}, black!55, line width=0.35pt, shorten >=1.5pt, shorten <=1pt]
  (lx) -- (axis cs:51510,4323.8);
\draw[-{Latex[length=1.2mm]}, black!55, line width=0.35pt, shorten >=1.5pt, shorten <=1pt]
  (lx) -- (axis cs:51510,139.7);

\nextgroupplot[
  xbar stacked, bar width=22pt, xmin=0, xmax=100,
  symbolic y coords={Joern,\enginename{}}, ytick=data,
  enlarge y limits=0.7,
  xlabel={\% of total time}, title={(c) Time breakdown},
  legend style={font=\scriptsize, at={(0.5,-0.28)}, anchor=north, legend columns=3,
    draw=black!35, fill=white, inner sep=2pt, column sep=4pt},
  legend image code/.code={\draw[#1] (0cm,-0.06cm) rectangle (0.18cm,0.10cm);},
]
\addplot[fill=phBuild, draw=phBuild!60!black, line width=0.3pt]
  coordinates {(69.8,Joern) (51.4,\enginename{})};
\addlegendentry{CPG build}
\addplot[fill=phLoad, draw=phLoad!60!black, line width=0.3pt]
  coordinates {(22.2,Joern) (4.5,\enginename{})};
\addlegendentry{CPG load}
\addplot[fill=phCall, draw=phCall!60!black, line width=0.3pt]
  coordinates {(0.0,Joern) (12.2,\enginename{})};
\addlegendentry{call graph}
\addplot[fill=phCand, draw=phCand!60!black, line width=0.3pt]
  coordinates {(7.2,Joern) (28.2,\enginename{})};
\addlegendentry{candidates}
\addplot[fill=phOther, draw=phOther!60!black, line width=0.3pt]
  coordinates {(0.8,Joern) (3.7,\enginename{})};
\addlegendentry{other}
\end{groupplot}
\end{tikzpicture}